\documentclass[
    ,final            
  ,numbers                 
  ]
  {aipproc}

\layoutstyle{6x9}

\begin{document}

\title{Cosmic Magnetic Fields:\\ Observations and Prospects}


\author{Rainer Beck}{address={Max-Planck-Institut f\"ur Radioastronomie,
Auf dem H\"ugel 69, 53121 Bonn, Germany} }

\begin{abstract}
Synchrotron emission, its polarization and its Faraday rotation at
radio frequencies of 0.2--10~GHz are powerful tools to study the
strength and structure of cosmic magnetic fields. Unpolarized
emission traces turbulent fields which are strongest in galactic
spiral arms and bars (20--30~$\mu$G) and in central starburst
regions (50--100~$\mu$G). Such fields are dynamically important,
e.g. they can drive gas inflows in central regions. Polarized
emission traces ordered fields which can be regular
(uni--directional) or anisotropic random (generated from isotropic
random fields by compression or shear). Ordered fields with spiral
patterns exist in grand-design, barred and flocculent galaxies, and
in central regions of starburst galaxies. The strongest ordered
(mostly regular) fields of 10--15~$\mu$G strength are generally
found in galactic interarm regions and follow the orientation of
adjacent gas spiral arms. Faraday rotation measures (RM) of the
diffuse polarized radio emission from the disks of several spiral
galaxies reveal large-scale patterns, which are signatures of
regular fields probably generated by a mean-field dynamo. Ordered
fields in interacting galaxies have asymmetric distributions and are
an excellent tracer of past interactions between galaxies or with
the intergalactic medium. Ordered magnetic fields are also observed
in radio halos around edge-on galaxies, out to large distances from
the plane, with X-shaped patterns. -- The strength of the total
magnetic field in our Milky Way is about 6~$\mu$G near the solar
radius, but several mG in dense clouds, pulsar wind nebulae, and
filaments near the Galactic Center. Diffuse polarized radio emission
and Faraday rotation data from pulsars and background sources show
spiral fields with large-scale reversals, but the overall field
structure in our Galaxy is still under debate. -- Diffuse radio
emission from the halos of galaxy clusters is mostly unpolarized
because intracluster magnetic fields are turbulent, while cluster
``relics'', probably shock fronts by cluster mergers, can have
degrees of polarization of up to 60\% and extents of up to 2~Mpc.
The IGM magnetic field strength is $\ge 3~10^{-16}$~G with a filling
factor of at least 60\%, derived from the combination of data from
the HESS and FERMI telescopes. -- Polarization observations with the
forthcoming large radio telescopes will open a new era in the
observation of cosmic magnetic fields and will help to understand
their origin. At low frequencies, LOFAR (10--250~MHz) will allow us
to map the structure of weak magnetic fields in the outer regions
and halos of galaxies and galaxy clusters. Small Faraday rotation
measures can also be best measured at low frequencies. Polarization
at higher frequencies (1--10~GHz), as observed with the EVLA,
MeerKAT, APERTIF and the SKA, will trace magnetic fields in the
disks and central regions of nearby galaxies in unprecedented
detail. The SKA pulsar survey will find many new pulsars; their RMs
will map the Milky Way's magnetic field with high precision. All-sky
surveys of Faraday rotation measures towards a dense grid of
polarized background sources with the SKA and its precursor
telescope ASKAP are dedicated to measure magnetic fields in distant
intervening galaxies, galaxy clusters and intergalactic filaments,
and will be used to model the overall structure and strength of the
magnetic field in the Milky Way. With the SKA, ordered fields in
distant galaxies and cluster relics can be measured to redshifts of
$z\simeq0.5$, turbulent fields in starburst galaxies or cluster
halos to $z\simeq3$ and regular fields in intervening galaxies
towards QSOs to $z\simeq5$.
\end{abstract}

\maketitle


\section{Introduction}
\label{intro}

Observations and modeling of cosmic magnetic fields revealed that
they are a major agent of the interstellar medium (ISM) and the
intracluster medium (ICM). They contribute significantly to the
total pressure which balances the gas disk of galaxies against
gravitation. They affect the dynamics of the turbulent ISM
\citep{avillez05} and the gas flows in spiral arms \citep{gomez02}.
The shock strength in spiral density waves is decreased and
structure formation is reduced in the presence of strong fields
\citep{dobbs08}. The interstellar fields are closely connected to
gas clouds. Magnetic fields stabilize gas clouds and reduce the
star-formation efficiency to the observed low values
\citep{price08,vazquez05}. On the other hand, magnetic fields are
essential for the onset of star formation as they enable the removal
of angular momentum from the protostellar cloud via ambipolar
diffusion \citep{heitsch04}. {\it MHD turbulence}\ distributes
energy from supernova explosions within the ISM \citep{subra98} and
drives field amplification and ordering via a {\it dynamo}\ (see
below).

Magnetic fields also control the density and distribution of cosmic
rays in the ISM. Cosmic rays accelerated in supernova remnants can
provide the pressure to drive a {\em galactic outflow}\ and buoyant
loops of magnetic fields via the {\em Parker instability}\
\citep{hanasz02}. Parker loops can in turn drive a dynamo
\citep{hanasz09}. Outflows from starburst galaxies in the early
Universe may have magnetized the intergalactic medium
\citep{kronberg99}. Understanding the interaction between the gas
and the magnetic field is a key to understand the physics of galaxy
disks and halos and the evolution of galaxies.

The detection of ultrahigh-energy cosmic rays (UHECRs) with the
AUGER observatory and the anisotropic distribution of their arrival
directions \citep{abreu10} calls for a proper model of particle
propagation. As UHECR particles are deflected by large-scale regular
fields and scattered by turbulent fields, the structure and the
extent of the fields in the disk and halo of the Milky Way need to
be known, but the present data do not allow safe conclusions
\citep{men08, noutsos09}.

Some fraction of massive galaxy clusters have radio halos which
require the generation of intracluster magnetic fields and the
acceleration of cosmic ray particles. The magnetic fields can affect
thermal conduction \citep{balbus08,parrish08} and hence the dynamics
and evolution of the intracluster medium.

In spite of our increasing knowledge, many important questions,
especially the origin and evolution of magnetic fields, their first
occurrence in young galaxies and galaxy clusters, or the strength
and structure of intergalactic fields remained unanswered.

\section{Origin of magnetic fields}

{\em Seed fields}\ may be ``primordial'', generated during a phase
transition in the early Universe \citep{caprini09,widrow02}, or
originate from the time of cosmological structure formation by the
Weibel instability \citep{lazar09}, or from injection by the first
stars or jets generated by the first black holes \citep{rees05}, or
from the Biermann mechanism in the first supernova remnants
\citep{hanayama05}.

The most promising mechanism to sustain magnetic fields in the
interstellar medium of galaxies is the dynamo \citep{beck96}. The
small-scale or {\em fluctuation dynamo}\ \citep{brand05} does not
need general rotation, only turbulent gas motions. It amplifies weak
seed fields to the energy density level of turbulence. A small-scale
dynamo in protogalaxies may have amplified seed fields to several
$\mu$G strength (the energy level of turbulence) within less than
$10^8$~yr \citep{schleicher10}. To explain the generation of
large-scale fields in galaxies, the mean-field dynamo has been
developed. It is based on turbulence, differential rotation and
helical gas flows ({\em $\alpha$--effect}), driven by supernova
explosions \citep{ferriere00,gressel08}. The mean-field dynamo in
galaxy disks predicts that within a few $10^9$~yr large-scale
regular fields are generated from $\mu$G turbulent fields
\citep{arshakian09}, forming spiral patterns ({\em modes}) with
different azimuthal symmetries in the disk and vertical symmetries
in the halo (see below). Global numerical models of galaxies
\citep{gissinger09,hanasz09} confirm the basic results of the
mean-field approximation.

The mean-field dynamo generates large-scale helicity with a non-zero
mean in each hemisphere. As total helicity is a conserved quantity,
the dynamo is quenched by the small-scale fields with opposite
helicity unless these are removed from the system
\citep{shukurov06}. Outflows are probably essential for effective
mean-field dynamo action.

In the intracluster medium, the seed fields ejected by galactic jets
or winds are probably amplified by a small-scale dynamo driven by
turbulent gas motions \citep{bertone06,subra06}.

\section{Tools to study magnetic fields}

Magnetic fields need illumination to be detectable. {\em Polarized
emission}\ at optical, infrared, submillimeter and radio wavelengths
holds the clue to measure magnetic fields in galaxies. Optical
linear polarization is a result of extinction by elongated dust
grains in the line of sight which are aligned in the interstellar
magnetic field (the {\em Davis-Greenstein effect}). The E--vector
runs parallel to the field. However, light can also be polarized by
scattering, a process unrelated to magnetic fields and hence a
contamination that is difficult to subtract from the diffuse
polarized emission from galaxies, e.g. in M~51 \citep{scarrott87}.
Optical polarization data of about 5500 selected stars in the Milky
Way yielded the orientation of the large-scale magnetic field near
the sun \citep{fosalba02}. Together with measurements of stellar
distances, a 3-D analysis of the magnetic field within about 5~kpc
from the sun is possible, but more data are needed.

Linearly polarized emission from elongated dust grains at infrared
and submillimeter wavelengths is not affected by polarized scattered
light. The B--vector is parallel to the magnetic field. The field
structure can be mapped in gas clouds of the Milky Way
\citep{tang09} and in galaxies, e.g. in the halo of M~82
\citep{greaves00}.

Most of what we know about interstellar magnetic fields comes
through the detection of radio waves. {\em Zeeman splitting}\ of
radio spectral lines directly measures the field strength in gas
clouds of the Milky Way \citep{crutcher10} and in starburst galaxies
\citep{robishaw08}. The intensity of {\em synchrotron emission}\ is
a measure of the density of cosmic-ray electrons in the relevant
energy range and of the strength of the total field component in the
sky plane. The assumption of energy equipartition between these two
components allows us to calculate the total magnetic field strength
from the synchrotron intensity \citep{beck+krause05}.

Linearly polarized synchrotron emission emerges from ordered fields
in the sky plane. As polarization ``vectors'' are ambiguous by
$180^\circ$, they cannot distinguish {\em regular (coherent)
fields}, defined to have a constant direction within the telescope
beam, from {\em anisotropic fields}, which are generated from
turbulent fields by compressing or shearing gas flows and frequently
reverse their direction within the telescope beam. Unpolarized
synchrotron emission indicates {\em turbulent (random) fields}\,
which have random directions in 3-D and have been amplified and
tangled by turbulent gas flows.

The intrinsic degree of linear polarization of synchrotron emission
is about 75\%. The observed degree of polarization is smaller due to
the contribution of unpolarized thermal emission, which may dominate
in star-forming regions, by {\em Faraday depolarization}\ along the
line of sight and across the beam \citep{burn66,sokoloff98}, and by
geometrical depolarization due to variations of the field
orientation within the beam.

The polarization vector is rotated in a magnetized thermal plasma by
{\em Faraday rotation}. If Faraday rotation is small (in galaxies
typically at wavelengths shorter than a few centimeters), the
observed $B$--vector gives the intrinsic field orientation in the
sky plane, so that the magnetic pattern can be mapped directly
\citep{beck05a}. As the rotation angle is sensitive to the sign of
the field direction, only regular fields give rise to Faraday
rotation, while anisotropic and random fields do not. Measurements
of the Faraday rotation from multi-wavelength observations allow to
determine the strength and direction of the regular field component
along the line of sight.

The rotation angle $\Delta\chi$ is proportional to the square of the
wavelength $\lambda^2$ and the {\em Faraday depth (FD)}, defined as
the line-of-sight integral over the product of the plasma density
and the strength of the field component along the line of sight. The
observable {\em rotation measure (RM)} is defined as
$RM=\Delta\chi/\Delta\lambda^2$. If the rotating region is located
in front of the emitting region ({\em Faraday screen}), $RM=FD$. In
case of a region with emission and rotation, $RM \approx FD /2$.
Several distinct emitting and rotating regions located along the
line of sight generate a spectrum of FD components. In such cases,
multi-channel spectro-polarimetric radio data are needed that can be
Fourier-transformed into Faraday space, called {\em RM Synthesis}\
\citep{brentjens05,frick11,heald09a,heald09b}. If the medium has a
simple magnetic structure, its 3-D structure can be determined ({\em
Faraday tomography}).

A {\em grid of RM measurements of polarized background sources}\ is
another tool to study magnetic field patterns in galaxies
\citep{stepanov08} and in the intracluster medium \citep{krause+09}.
A large number of sources is required to recognize the field
patterns, to separate the Galactic foreground contribution and to
account for intrinsic RMs of the background sources.

The combination of observations of diffuse polarized emission and RM
can be used as a test of {\em magnetic helicity}\
\citep{oppermann10}.

\section{Total fields in galaxies}

\begin{figure*}[t]
\includegraphics[bb = 47 47 522 241,width=12cm,clip=]{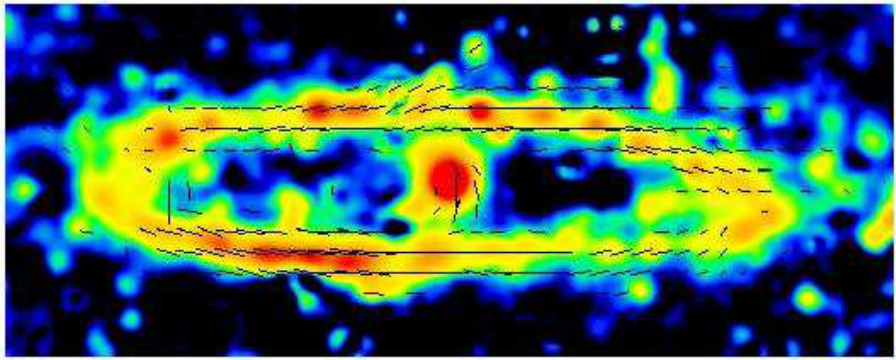}
\caption{Total radio intensity (colors) and B--vectors (corrected
for Faraday rotation) in the Andromeda galaxy (M~31), observed at
6~cm with the Effelsberg telescope \citep{berk03}.} \label{fig:m31}
\end{figure*}

The typical average {\em equipartition strength}\ of the total
magnetic field \citep{beck+krause05} in spiral galaxies is about
$10~\mu$G, assuming energy equipartition between cosmic rays and
magnetic fields. Radio-faint galaxies like M~31 (Fig.~\ref{fig:m31})
and M~33 \citep{taba08}, our Milky Way's neighbors, have weaker
total magnetic fields (about $6~\mu$G), while gas-rich spiral
galaxies with high star-formation rates, like M~51 and NGC~6946
(Fig.~\ref{fig:m51}), have total field strengths of $20-30~\mu$G in
their spiral arms. The strongest total fields of $50-100~\mu$G are
found in starburst galaxies, like M~82 \citep{klein88} and the
``Antennae'' NGC~4038/9 \citep{chyzy04}, and in nuclear starburst
regions, like in the centers of NGC~1097 and other barred galaxies
\citep{beck05b}.

If energy losses of cosmic-ray electrons are significant, especially
in starburst regions or massive spiral arms, the equipartition
values are lower limits \citep{beck+krause05} and are probably
underestimated in starburst galaxies by a factor of a few
\citep{thompson06}. Field strengths of $0.5-18$~mG were detected in
starburst galaxies by the Zeeman effect in the OH megamaser emission
line at 18~cm \citep{robishaw08}. These values refer to highly
compressed gas clouds and are not typical for the diffuse
interstellar medium.

\begin{figure*}[t]
\includegraphics[width=10cm]{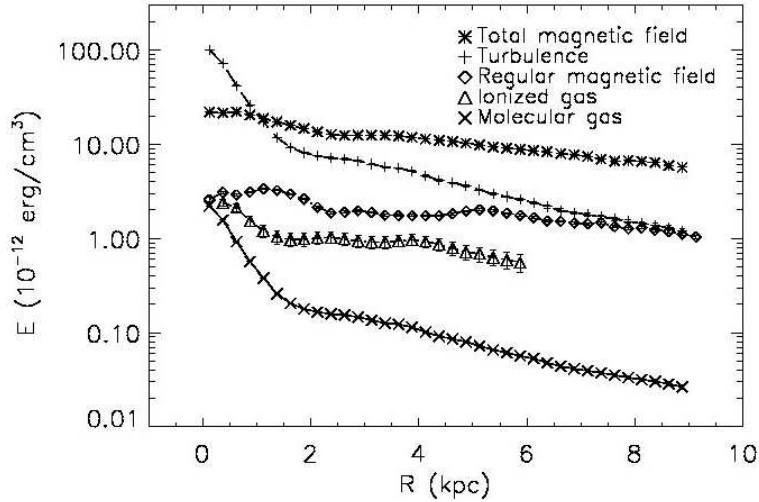}
\caption{Radial variation of the energy densities in NGC~6946: total
magnetic field $E_B$ ($B_t^2/8\,\pi$), regular magnetic field
($B_{reg}^2/8\,\pi$), turbulent motion of the neutral gas $E_{turb}$
($0.5\,\rho_n\,v_{turb}^2$, where $v_{turb}\approx7$~km/s), thermal
energy of the ionized gas $E_{th}$ ($0.5\,n_e\,k\,T_e$) and thermal
energy of the molecular gas $E_n$ ($0.5\,\rho_n\,k\,T_n$),
determined from observations of synchrotron and thermal radio
continuum and the CO and HI line emissions \citep{beck07a}.}
\label{fig:n6946_energies}
\end{figure*}

The relative importance of various competing forces in the
interstellar medium are estimated by comparing {\em energy
densities}. The mean energy densities of the total (mostly
turbulent) magnetic field and the cosmic rays in NGC~6946
(Fig.~\ref{fig:n6946_energies}) and M~33 are
$\simeq10^{-11}$~erg~cm$^{-3}$ and $\simeq10^{-12}$~erg~cm$^{-3}$,
respectively \citep{beck07a,taba08}, similar to that of the
turbulent gas motions in the star-forming disk, about 10 times
larger than that of the ionized gas ({\em low-beta plasma}).
Magnetic fields are dynamically important. The total magnetic energy
density may even dominate in the outer galaxy where the
equipartition field strength is an underestimate due to energy
losses of the cosmic-ray electrons. The energy density of the
regular magnetic field decreases even more slowly than that of the
total field, possibly because the mean-field dynamo still
efficiently operates in the outer disk. Although the star-formation
activity is low here, the magneto-rotational instability (MRI) may
serve as the source of turbulence required for dynamo action
\citep{sellwood99}.


The integrated luminosity of the total radio continuum emission at
centimeter wavelengths (frequencies of a few GHz), which is mostly
of nonthermal synchrotron origin, and the far-infrared (FIR)
luminosity of star-forming galaxies are tightly correlated. This
correlation is one of the tightest correlations known in astronomy.
It extends over five orders of magnitude \citep{bell03} and is valid
in starburst galaxies to redshifts of at least 3 \citep{seymour08}.
Hence the total radio emission can serve as a tracer of magnetic
fields and of star formation out to large distances. The correlation
requires that total (mostly turbulent) magnetic fields and star
formation are connected, so that the field strength exceeds several
100~$\mu$G in distant galaxies \citep{murphy09}. The tightness needs
multiple feedback mechanisms which are not yet understood
\citep{lacki10}.

The total radio and far-infrared (FIR) or mid-IR (MIR) intensities
are also highly correlated within galaxies. The exponent of the
correlation in M~51 was found to be different in the central region,
spiral arms and interarm regions \citep{dumas11}. The
radio--infrared correlation can be presented as a correlation
between turbulent field strength and star-formation rate
\citep{chyzy08a}. In contrast, the ordered field is either
uncorrelated with the star-formation rate, or anticorrelated in
galaxies where the ordered field is strongest in interarm regions
with low star formation (Fig.~\ref{fig:m51}). A wavelet
cross-correlation analysis for M~33 showed that the radio--FIR
correlation holds at all scales down to 1~kpc \citep{taba07}. The
correlation in the Large Magellanic Cloud (LMC) breaks down below
scales of about 50~pc \citep{hughes06}, probably due to the
diffusion of cosmic-ray electrons.

\section{Ordered magnetic fields in galaxies}

\begin{figure*}[t]
\vspace*{2mm}
\begin{minipage}[t]{7.3cm}
\includegraphics[bb = 47 47 522 440,width=7.1cm,clip=]{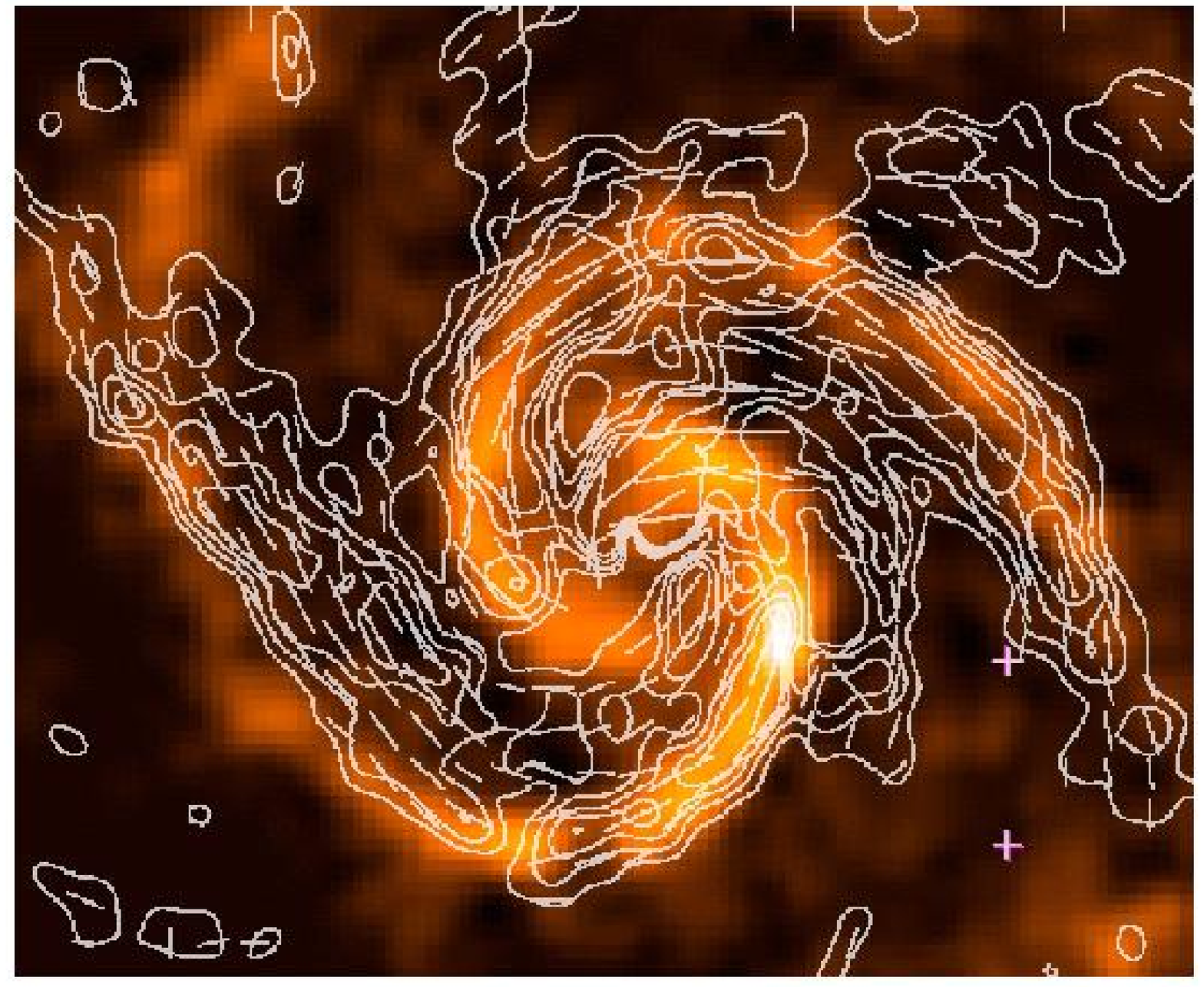}
\end{minipage}\hfill
\begin{minipage}[t]{7.5cm}
\includegraphics[bb = 20 20 592 489,width=7.1cm,clip=]{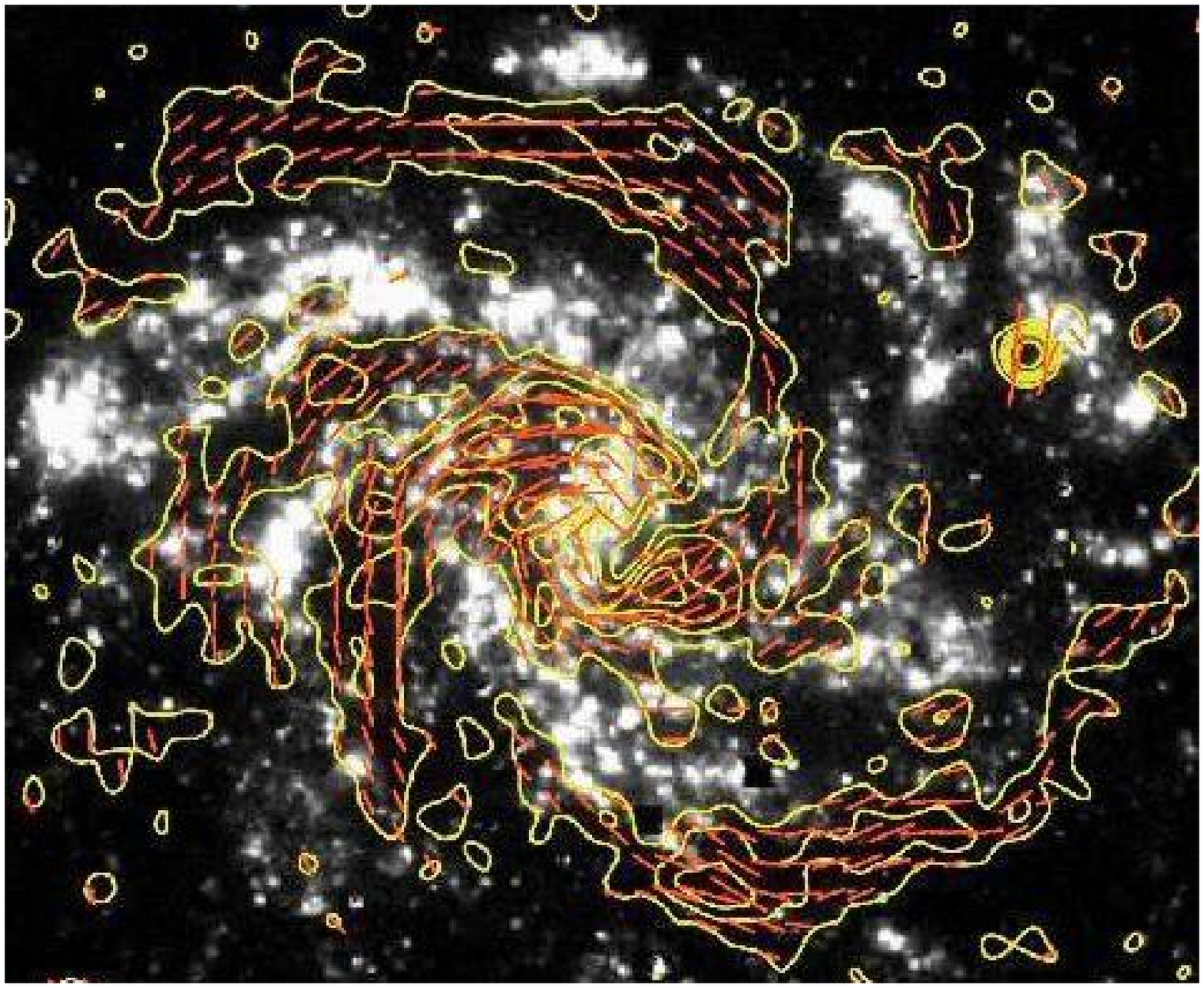}
\caption{{\bf Left:} Polarized radio intensity (contours) and
B--vectors in the inner 3'x4' of M~51, combined from observations at
6~cm wavelength with the VLA and Effelsberg telescopes at 8''
resolution \citep{fletcher11}, overlaid onto an image of the
molecular CO(1-0) line emission \citep{helfer03}. {\bf Right:}
Polarized radio intensity (contours) and B--vectors of NCC~6946,
combined from observations at 6~cm wavelength with the VLA and
Effelsberg telescopes and smoothed to 15'' resolution
\citep{beck07a}, overlaid onto an H$\alpha$ image from Anne Ferguson
(Copyright: MPIfR Bonn; graphics: \textit{Sterne und Weltraum}).}
\label{fig:m51}
\end{minipage}
\end{figure*}

Ordered (regular and/or anisotropic) field traced by polarized
synchrotron emission form spiral patterns in almost every galaxy
\citep{beck05a}, even in ring galaxies \citep{chyzy08b}, in
flocculent galaxies without massive spiral arms \citep{soida02} and
in the central regions of galaxies and in circum-nuclear gas rings
of barred galaxies \citep{beck05b}. Ordered fields are generally
strongest (10--15~$\mu$G) in the regions {\em between}\ the optical
spiral arms and oriented parallel to the adjacent spiral arms, in
some galaxies forming {\em magnetic arms}, like in NGC~6946
(Fig.~\ref{fig:m51} right), with exceptionally high degrees of
polarization (up to 50\%). These are probably generated by a
large-scale dynamo (see below). In galaxies with strong density
waves like M~51 \citep{fletcher11} enhanced ordered (anisotropic)
fields occur at the inner edges of the inner optical arms and in the
interarm regions (Fig.~\ref{fig:m51} left).

The observed {\em spiral magnetic patterns}\ with significant pitch
angles (in the range $10^\circ-40^\circ$) indicate a general
decoupling between magnetic fields and the gas flow, as predicted by
mean-field dynamo action. There is no other model to explain the
magnetic spiral patterns in many types of galaxies.

The typical degree of radio polarization within the spiral arms is
only a few \%; hence the field in the spiral arms must be mostly
tangled or randomly oriented within the telescope beam, the width of
which corresponds to a few 100~pc. Turbulent fields in spiral arms
are probably generated by turbulent gas motions related to star
formation activity or by a small-scale dynamo \citep{subra98}.

\begin{figure*}[t]
\vspace*{2mm}
\begin{minipage}[t]{7.5cm}
\includegraphics[bb = 64 176 552 634,width=7cm,clip=]{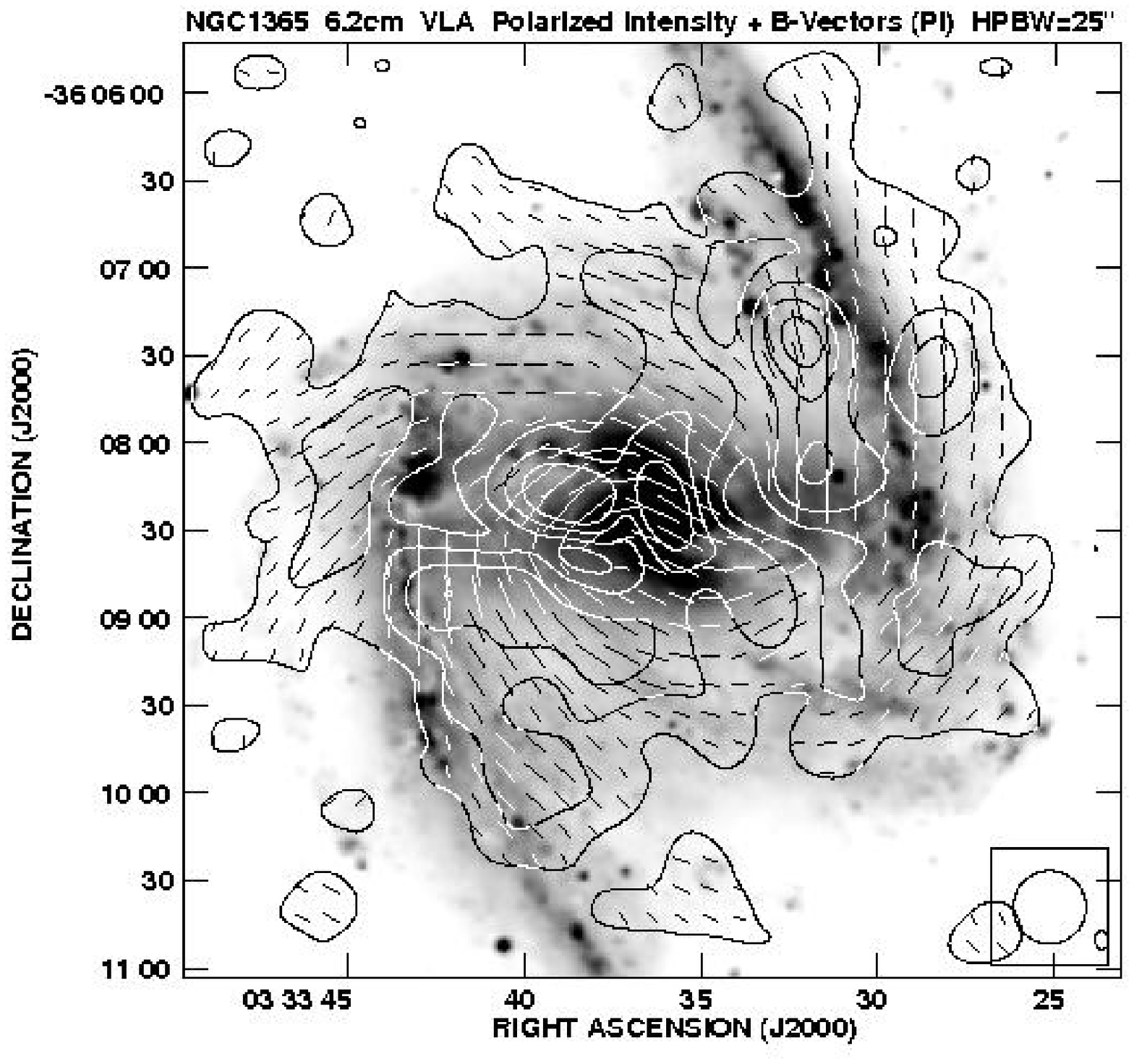}
\end{minipage}\hfill
\begin{minipage}[t]{7.5cm}
\includegraphics[bb = 47 47 522 505,width=6.7cm,clip=]{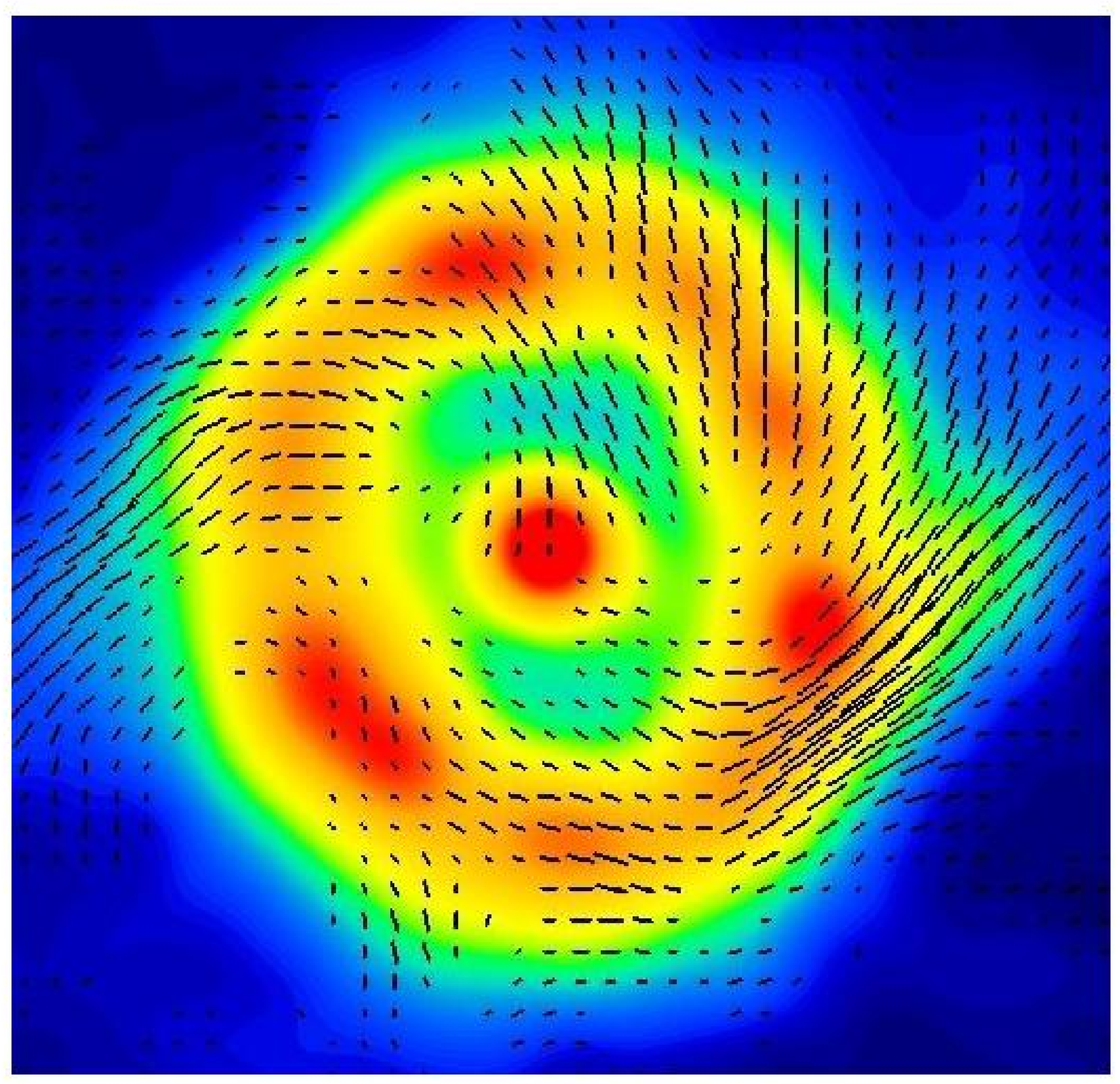}
\caption{{\bf Left:} Total radio intensity and B--vectors of the
barred galaxy NGC~1365, observed at 6.2~cm wavelength with the VLA,
smoothed to 25'' resolution and overlaid onto an optical image
\citep{beck05b}. {\bf Right:} Total radio intensity and B--vectors
in the circumnuclear ring of the barred galaxy NGC~1097, observed at
3.5~cm wavelength with the VLA at 3'' resolution \citep{beck05b}
(Copyright: MPIfR Bonn).} \label{fig:n1097}
\end{minipage}
\end{figure*}

In galaxies with massive {\em bars}\ the field lines follow the gas
flow (Fig.~\ref{fig:n1097} left). As the gas rotates faster than the
bar pattern of a galaxy, a shock occurs in the cold gas that has a
small sound speed, while the flow of warm, diffuse gas is only
slightly compressed but sheared. The ordered field is also hardly
compressed. It is probably coupled to the diffuse gas and strong
enough to affect its flow \citep{beck05b}.
The polarization pattern in barred galaxies can be used as a tracer
of shearing gas flows in the sky plane and hence complements
spectroscopic measurements of radial velocities.

The {\em central regions}\ of barred galaxies are often sites of
ongoing intense star formation and strong magnetic fields that can
affect the gas flow. NGC~1097 hosts a bright ring with about
1.5~kiloparsec diameter and an active nucleus in its center
(Fig.~\ref{fig:n1097} right). The ordered field in the ring has a
spiral pattern and extends towards the nucleus. The orientation of
the innermost spiral field agrees with that of the spiral dust
filaments visible on optical images. Magnetic stress in the
circumnuclear ring due to the strong total magnetic field (about
50~$\mu$G) can drive gas inflow \citep{balbus98} at a rate of
several M$_{\circ}$/yr, which is sufficient to fuel the activity of
the nucleus \citep{beck05b}.

\begin{figure*}[t]
\vspace*{2mm}
\begin{minipage}[t]{7.5cm}
\includegraphics[bb = 32 32 482 448,width=6.5cm,clip=]{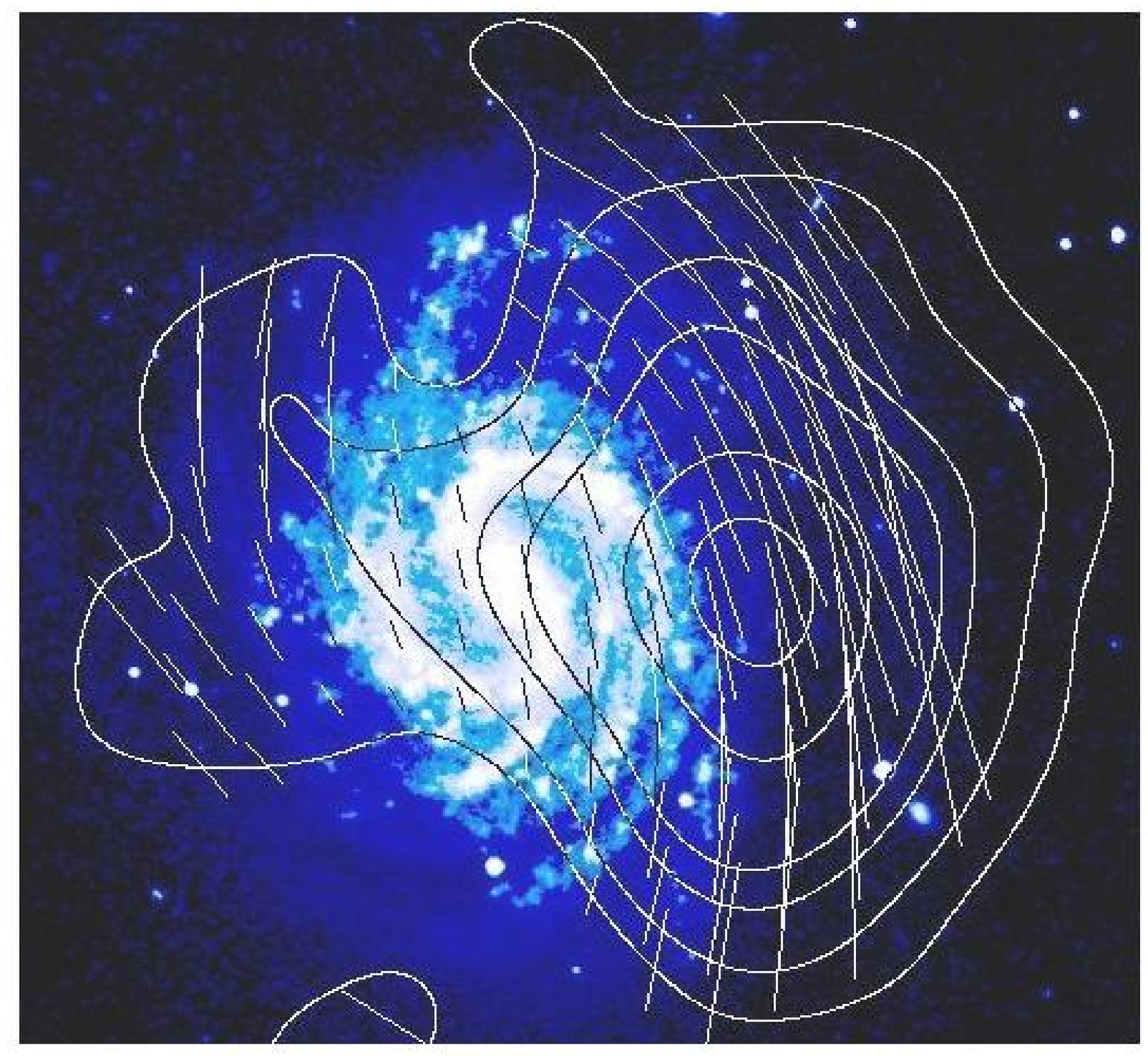}
\end{minipage}\hfill
\begin{minipage}[t]{7cm}
\includegraphics[bb = 32 32 443 515,width=6cm,clip=]{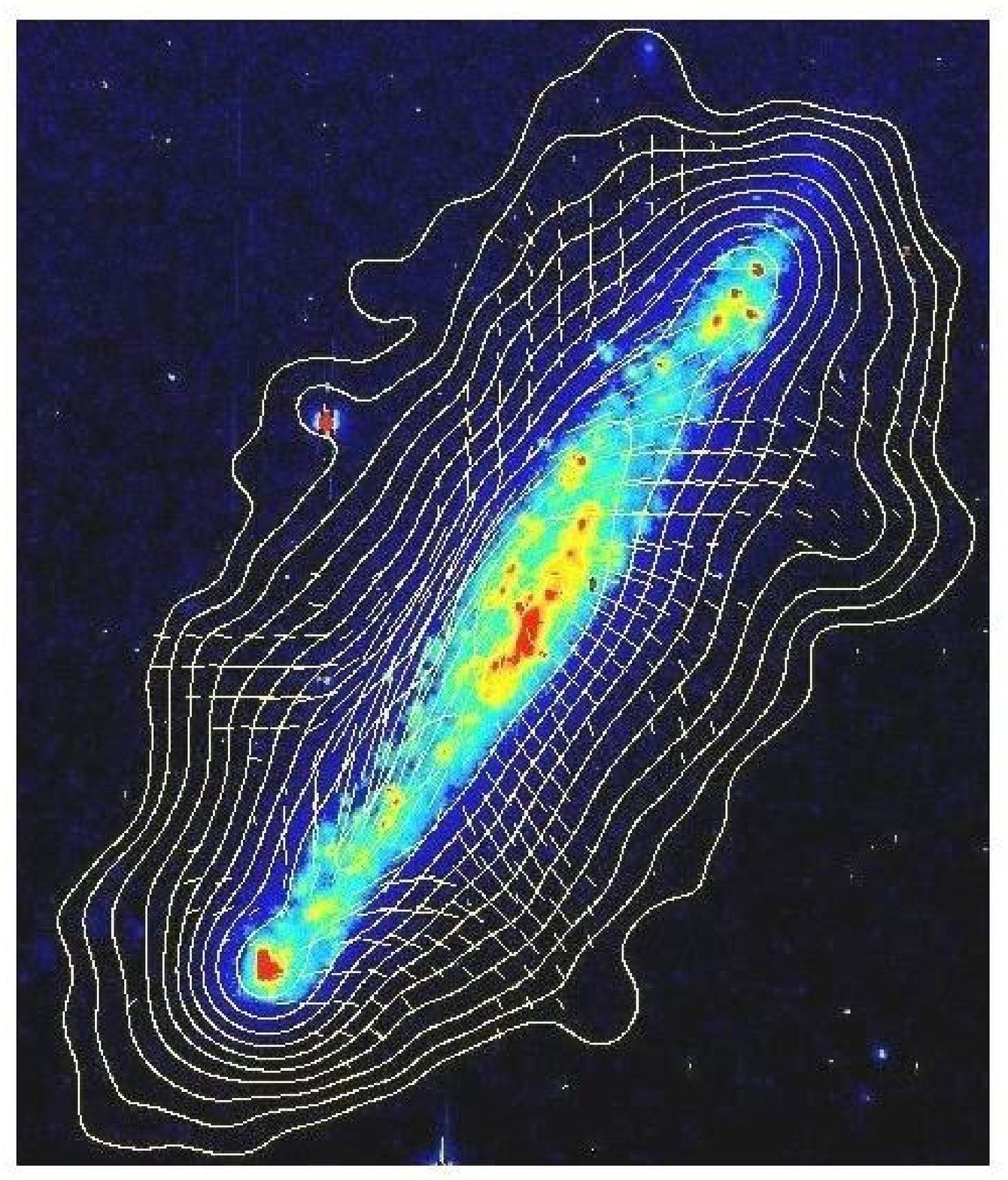}
\caption{{\bf Left:} Polarized radio intensity (contours) and
B--vectors of the spiral galaxy NGC~4535 in the Virgo cluster,
observed at 6.3~cm with the Effelsberg telescope \citep{wez07}. The
background optical image is from the Digital Sky Survey. {\bf
Right:} Total radio intensity and B--vectors of the edge-on galaxy
NGC~5775, observed at 6.2~cm wavelength with the VLA at 17''
resolution \citep{tuell00}.} \label{fig:n5775}
\end{minipage}
\end{figure*}

{\em Interaction}\ with a dense intergalactic medium also imprints
unique signatures onto magnetic fields and thus the radio emission.
The Virgo cluster is a location of especially strong interaction
effects (Fig.~\ref{fig:n5775} left), and almost all cluster galaxies
observed so far show asymmetries of their polarized emission because
the outer magnetic fields were compressed by ram pressure or
shearing gas flows \citep{vollmer07,wez07}. Ordered fields are an
excellent tracer of past interactions between galaxies or with the
intergalactic medium, or sweeping-up of the intracluster field
\citep{pfrommer10}.

{\em Flocculent}\ galaxies have disks but no prominent spiral arms.
Nevertheless, spiral magnetic patterns exist in all flocculent
galaxies observed so far, indicative that the mean-field dynamo
works independently of density waves. Ordered magnetic fields with
strengths similar to those in grand-design spiral galaxies have been
detected in the flocculent galaxies M~33 \citep{taba08}, NGC~3521
and NGC~5055 \citep{knapik00}, and in NGC~4414 \citep{soida02}. The
mean degree of polarization (corrected for the differences in
spatial resolution) is similar between grand-design and flocculent
galaxies \citep{knapik00}.

Radio continuum maps of {\em irregular}, slowly rotating galaxies
may reveal strong total equipartition magnetic fields, e.g. in the
Magellanic-type galaxy NGC~4449 where a fraction of the field is
ordered with about 7~$\mu$G strength and a spiral pattern
\citep{chyzy00}. Faraday rotation shows that this ordered field is
mostly regular and the mean-field dynamo is operating in this galaxy.
Dwarf irregular galaxies with almost chaotic rotation do not have any
regular fields and only spots of faint polarized emission
\citep{chyzy03}. The
turbulent field strengths in starburst dwarfs are comparable to
those in large spiral galaxies, e.g. in NGC~1569 \citep{kepley10},
probably driven by a small-scale dynamo.

Nearby galaxies seen {\em edge-on}\ generally show a disk-parallel
field near the disk plane \citep{dumke95}. As a result, polarized
emission can also be detected from distant, unresolved galaxies if
the inclination is larger than about $20^{\circ}$ \citep{stil09}.
This opens a new method to search for ordered fields in distant
galaxies. High-sensitivity radio polarization observations of
edge-on galaxies like NGC~253 \citep{heesen09b}, NGC~891
\citep{krause09} and NGC~5775 (Fig.~\ref{fig:n5775} right) revealed
vertical field components in the halo forming an X-shaped pattern
which may be related to dynamo action \citep{moss10} or outflows
\citep{vecchia08}.

The stronger magnetic field in the central regions leads to larger
synchrotron loss, leading to the ``dumbbell'' shape of many radio
halos, e.g. in NGC~253 \citep{heesen09a}, which is in contrast to
its almost spherical X-ray halo \citep{pietsch00}. From the radio
scale heights at several frequencies and the corresponding electron
lifetimes (due to synchrotron, IC and adiabatic losses) a transport
speed of about 300~km/s was measured for the halo of NGC~253
\citep{heesen09a}. Similar radio scale heights of about 2~kpc of
most edge-on galaxies observed so far, in spite of the different
field strengths, indicates that the outflow speed increases with the
average field strength and the star-formation rate \citep{krause09}.

\begin{table}
\begin{tabular}{lll}
\hline
  \tablehead{1}{l}{b}{Galaxy type}
  & \tablehead{1}{l}{b}{Magnetic field structure}
  & \tablehead{1}{l}{b}{Regular field (dynamo)}\\
\hline
Sc galaxy with strong & Spiral field at inner edge and in & Weak or moderate\\
\,\,\, density wave & \,\,\, interarm regions, turbulent field in arms\\
Sb or Sc galaxy with weak & Spiral field in interarm regions, & Strong\\
\,\,\, or moderate density wave & \,\,\, turbulent field in arms\\
Barred Sc galaxy & Ordered + turbulent field along bar, & Weak\\
 & spiral field outside bar \\
 Flocculent Sc or Sd galaxy & Spiral + turbulent field in disk & Weak\\
Irregular galaxy & Turbulent field in star-forming regions & Not detected\\
 & + segments of ordered field\\
Starburst dwarf galaxy & Turbulent field in star-forming regions & Not detected\\
Spheroidal dwarf galaxy & Not detected & Not detected\\
Sa galaxy & Ordered + turbulent fields & Not detected\\
S0 galaxy & Not detected & Not detected\\
E galaxy (non-active nucleus) & Not detected & Not detected\\
\hline
\end{tabular}
\caption{Typical field structures in galaxies} \label{tab:summary}
\end{table}

In the exceptionally large radio halos around the irregular and
interacting galaxies M~82 \citep{reuter94} and NGC~4631
\citep{golla94} a few magnetic spurs could be resolved, connected to
star-forming regions. These observations support the idea of a
strong galactic outflow that is driven by regions of star formation
in the inner disk.

{\em Early-type galaxies}\ (Sa, S0) and elliptical galaxies without
an active nucleus have very little star formation and hence do not
produce many cosmic rays emitting synchrotron emission. The only
deep observation of a Sa galaxy, M~104 with a prominent dust ring,
revealed weak, ordered magnetic fields \citep{krause06}. Large-scale
regular magnetic fields may exist in differentially rotating
galaxies even without any star formation because turbulence can be
generated by the magneto-rotational instability (MRI)
\citep{sellwood99}. Their detection may become possible via RM grids
of background sources with future radio telescopes.

\section{Faraday rotation and dynamos in galaxies}

Spiral magnetic patterns in galaxies can be generated by several
mechanisms, like compression at the inner edge of spiral arms, by
shear in interarm regions, or by dynamo action. For a distinction,
{\em Faraday rotation measures}\ (RM) need to be measured from the
diffuse polarized emission of the galaxy or from RM data of
polarized background sources \citep{stepanov08}. Large-scale RM
patterns are a clear signature a mean-field dynamo. Several dynamo
modes can be superimposed, so that a Fourier analysis is needed
\citep{krause90}. The resolution of present-day observations is
sufficient to identify 2--3 modes.

\begin{figure*}[t]
\includegraphics[bb = 51 51 519 232,width=12cm,clip=]{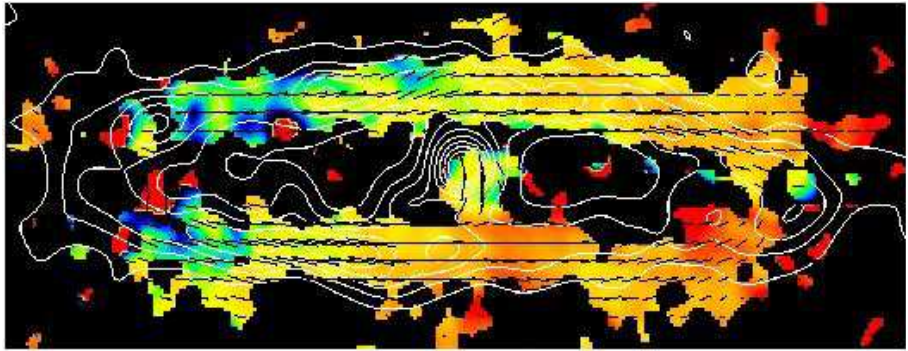}
\caption{Total radio intensity (contours) and Faraday Rotation
Measures (RM) in the Andromeda galaxy (M~31), determined from
observations at 6~cm and 11~cm with the Effelsberg telescope
\citep{berk03}. The intrinsic RM varies between about
-80~rad~m$^{-2}$ (left side of the major axis) and about
+80~rad~m$^{-2}$ (right side of the major axis).}
\label{fig:m31rm}
\end{figure*}

The disks of about a dozen of nearby spiral galaxies reveal
large-scale RM patterns. The Andromeda galaxy M~31
(Fig.~\ref{fig:m31rm}) is the prototype of a dynamo-generated
axisymmetric spiral disk field \citep{fletcher04} which extends to
25~kpc distance from the center \citep{han98}. Other candidates for
a dominating axisymmetric disk field (dynamo mode $m = 0$) are the
nearby spiral IC~342 \citep{krause89a} and the irregular Large
Magellanic Cloud (LMC) \citep{gaensler05}. Dominating bisymmetric
spiral fields (dynamo mode $m = 1$) are rare, with M~81 as the only
known case \citep{krause89b}. Faraday rotation in the magnetic arms
of NGC~6946 (Fig.~\ref{fig:m51} right) and in other similar galaxies
can be described by a superposition of two azimuthal dynamo modes
($m = 0$ and $m = 2$) with about equal amplitudes where the
quadrisymmetric spiral mode is phase shifted with respect to the
density wave \citep{beck07a}.

The spiral pattern of magnetic fields cannot be solely the result of
mean-field dynamo action. If the beautiful spiral pattern of M~51
seen in radio polarization \citep{fletcher11} were only due to a
regular field, its line-of sight component should generate a
conspicuous large-scale pattern in Faraday rotation, which is not
observed. This means that a large amount of the ordered field is
{\em anisotropic}\ and probably generated by compression and shear
of the non-axisymmetric gas flows in the density-wave potential. The
anisotropic field is strongest at the positions of the prominent
dust lanes on the inner edge of the inner gas spiral arms, due to
compression of turbulent fields in the density-wave shock. A weak
regular field (dynamo modes $m = 0$ and $m = 1$) also exists in the
disk of M~51 \citep{fletcher11}.

In many other observed galaxy disks no clear patterns of Faraday
rotation were found. Either several dynamo modes are superimposed
and cannot be distinguished with the limited sensitivity and
resolution of present-day telescopes, or the timescale for the
generation of large-scale modes is longer than the galaxy's lifetime
\citep{arshakian09}.


While the azimuthal symmetry of the magnetic field is known for many
galaxies, the vertical symmetry (even or odd) is much harder to
determine. The RM patterns of even and odd modes are similar in
mildly inclined galaxies. The field of odd modes reverses its sign
above and below the galactic plane. The symmetry type becomes only
visible in strongly inclined galaxies, as the RM sign above and
below the plane. Faraday RM data of NGC~253 indicate an
even-symmetry field \citep{heesen09b}. Indirect evidence for
dominating even fields \citep{braun10} comes from the asymmetry of
polarized emission along the major axis observed at 1.4~GHz in many
galaxies \citep{heald09b}.

{\em Large-scale field reversals}\ at certain radial distances from
a galaxy's center, like those in the Milky Way (see below), have not
been detected in spiral galaxies so far, although high-resolution RM
maps of Faraday rotation are available for many spiral galaxies. In
the barred galaxy NGC~7479, where a jet serves as a bright polarized
background and high-resolution observations were possible with high
signal-to-noise ratio, several reversals on 1--2~kpc scale were
detected in the foreground disk of the galaxy \citep{laine08}.

\section{Magnetic fields in the Milky Way}

The detection of ultrahigh-energy cosmic rays (UHECRs) with the
AUGER observatory and the anisotropic distribution of their arrival
directions \citep{abreu10} calls for a proper model of particle
propagation. As UHECR particles are deflected by large-scale regular
fields and scattered by turbulent fields, the structure and extent
of the fields in the disk and halo of the Milky Way need to be
known.

Optical polarization data of about 5500 selected stars in the Milky
Way yielded the orientation of the large-scale magnetic field near
the sun \citep{fosalba02}, which is mostly parallel to the Galactic
plane and oriented along the local spiral arm.

\begin{figure*}[t]
\vspace*{2mm}
\begin{minipage}[t]{9.5cm}
\includegraphics[bb = 20 20 592 308,width=9.5cm,clip=]{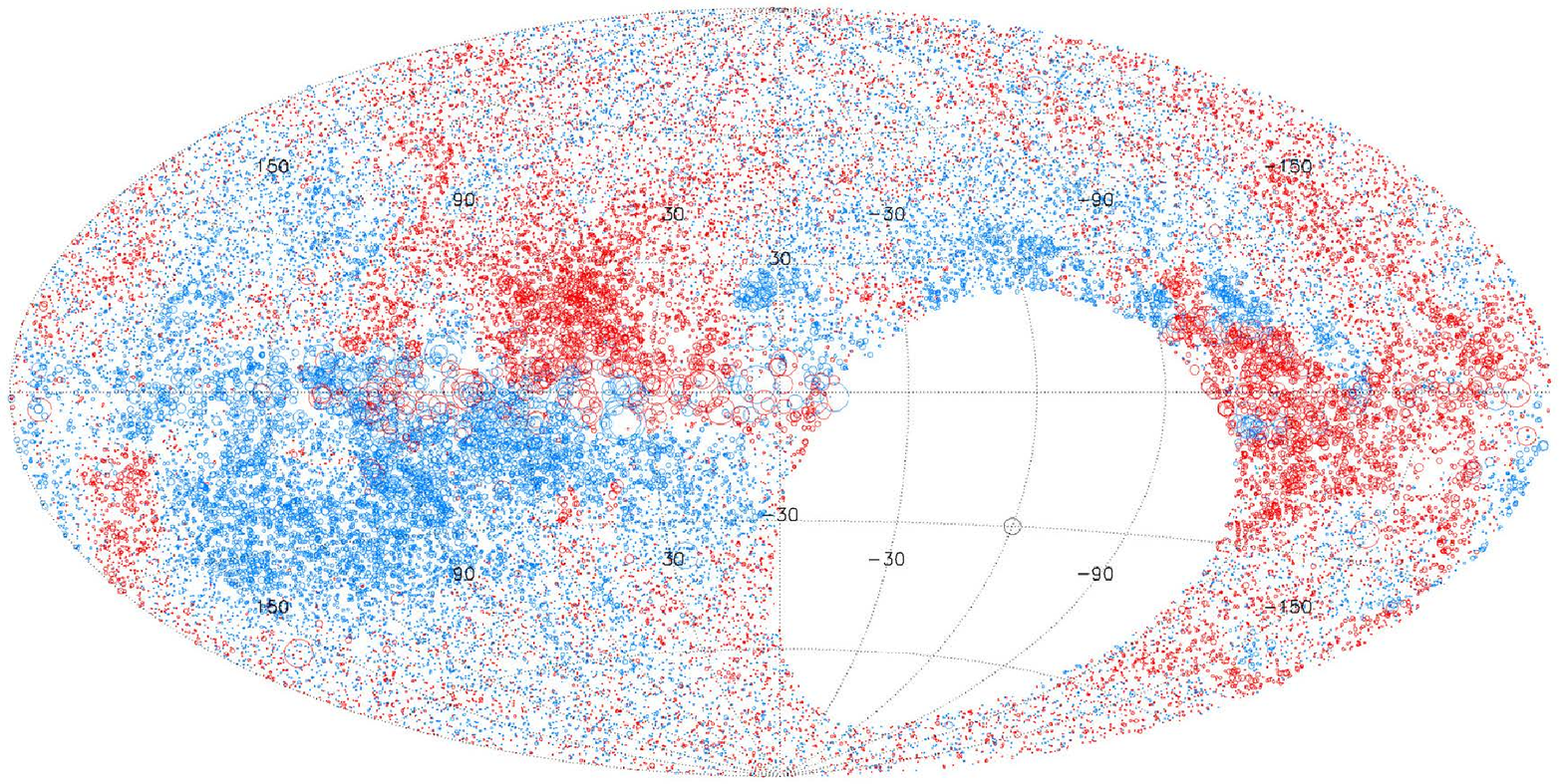}
\end{minipage}\hfill
\begin{minipage}[t]{5.3cm}
\includegraphics[bb = 20 20 582 590,width=5.1cm,clip=]{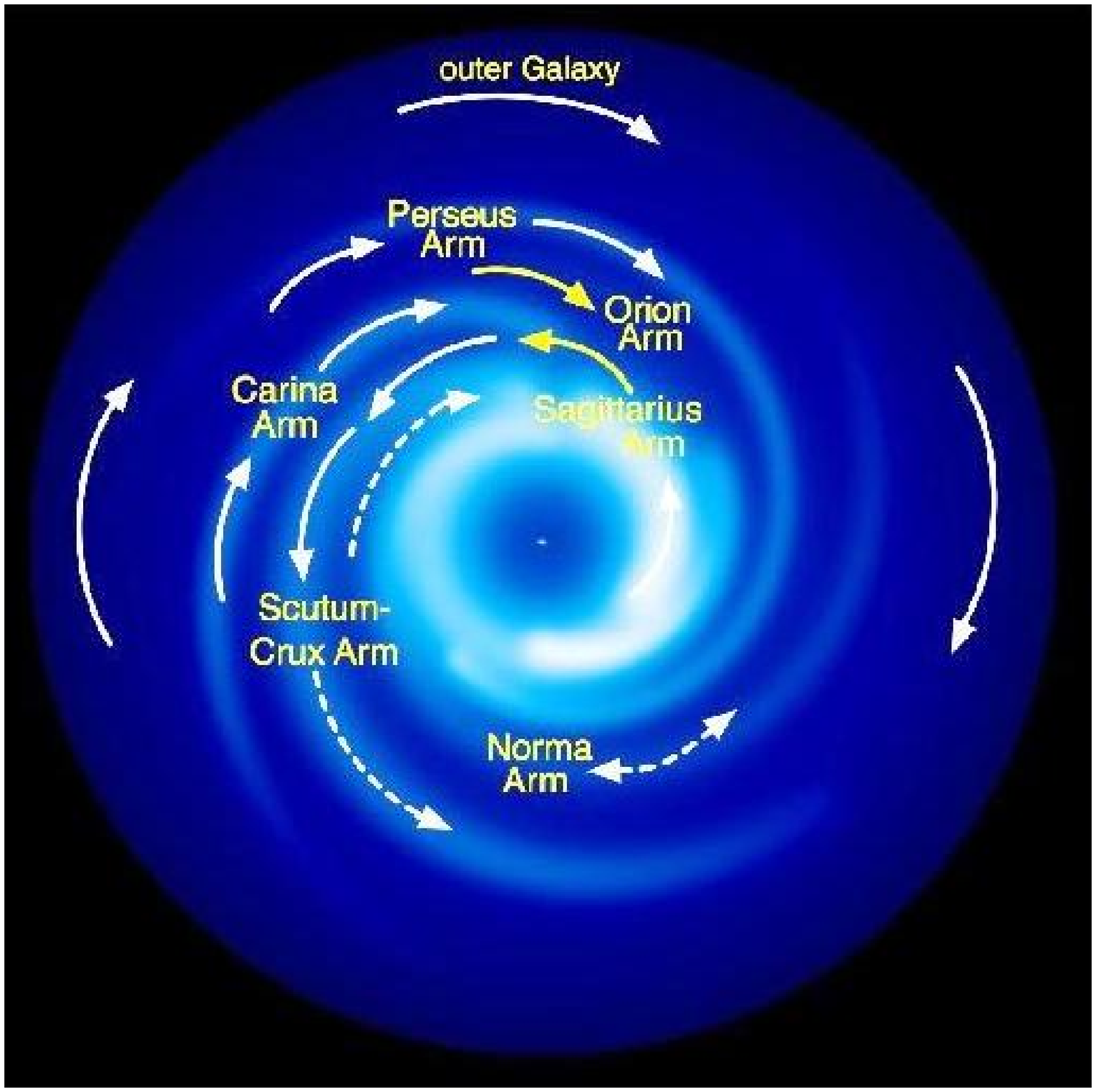}
\caption{{\bf Left:} All-sky map of rotation measures in the Milky
Way, generated from the data of 37,543 polarized extragalactic
sources from the VLA NVSS survey. Red circles: positive RM, blue
circles: negative RM; the circle size scales with $|RM|$
\citep{taylor09}. {\bf Right:} Model of the magnetic field in the
Milky Way, derived from Faraday rotation measures of pulsars and
extragalactic sources. Generally accepted results are indicated by
yellow vectors, while white vectors are not fully confirmed (from
Brown, priv. comm.).} \label{fig:Galaxy}
\end{minipage}
\end{figure*}

Surveys of the total synchrotron emission from the Milky Way yield
equipartition strengths of the total field of 6~$\mu$G near the sun
and about 10~$\mu$G in the inner Galaxy (Berkhuijsen, in
\citep{wielebinski05}), consistent with Zeeman splitting data of
low-density gas clouds \citep{crutcher10}. In dense dust clouds
field strengths of about $100~\mu$G were measured from submillimeter
polarimetry \citep{crutcher04}. Zeeman splitting of OH maser lines
from dense clouds yield field strengths of a few mG \citep{fish03}.
Milligauss fields were also found in pulsar wind nebulae from the
break in the synchrotron spectrum \citep{kothes08}.

In the nonthermal filaments near the Galactic Center the field
strength is several 100~$\mu$G \citep{ferriere09,reich94,yusef96}.
Their vertical orientation could be part of a poloidal field
\citep{ferriere09}, which is hard to observe in external galaxies
due to insufficient resolution. The break in the synchrotron
spectrum requires that the field near the Galactic Centre is at
least 50~$\mu$G on 400~pc scale \citep{crocker10}.

The all-sky maps of polarized synchrotron emission at 1.4~GHz from
the Milky Way from DRAO and Villa Elisa and at 22.8~GHz from WMAP
and the Effelsberg RM survey of polarized extragalactic sources were
used to model the regular Galactic field \citep{sun08,sun10}. One
large-scale {\em field reversal}\ is required at about 1--2~kpc from
the sun towards the Milky Way's center, which is also supported by
pulsar RMs \citep{frick01,nota10} and RMs from extragalactic sources
near the Galactic plane \citep{vaneck11} (Fig.~\ref{fig:Galaxy}
right). More large-scale reversals may exist \citep{han06}.

A satisfying explanation for the reversals in the Milky Way is still
lacking. Nothing similar has been detected in external galaxies so
far. The reversals may be restricted to a thin layer near to the
plane and hence are hardly visible in the average RM data of
external galaxies along the line of sight. Secondly, the reversals
in the Milky Way may be of limited azimuthal extent, and then are
difficult to observe in external galaxies with the resolution of
present-day telescopes. Thirdly, the reversals in the Milky Way may
be part of a disturbed field structure, e.g. due to interaction with
the Magellanic clouds.

The signs of RMs of extragalactic sources and of pulsars at Galactic
longitudes l=90$^\circ-270^\circ$ are the same above and below the
plane (Fig.~\ref{fig:Galaxy} left): the local magnetic field is
symmetric, while the RM signs towards the inner Galaxy
(l=270$^\circ-90^\circ$) are {\em opposite}\ above and below the
plane. This can be assigned to an antisymmetric halo field
\citep{sun10} or to deviations of the local field
\citep{wolleben10}. In conclusion, the overall structure of the
regular field in the disk of the Milky Way is not known yet
\citep{men08,noutsos09}. A larger sample of pulsar and extragalactic
RM data is needed.

While the large-scale field is much more difficult to measure in the
Milky Way than in external galaxies, Galactic observations can trace
magnetic structures to much smaller scales \citep{reich06}.
Polarization surveys at 1.4~GHz \citep{landecker10,wolleben06} and
at lower frequencies \citep{haverkorn04,schnitzeler09} reveal a
wealth of structures on parsec and sub-parsec scales: filaments,
canals, lenses and rings. They appear only in the maps of polarized
intensity, but not in total intensity. Some of these are artifacts
due to strong depolarization of background emission in a foreground
Faraday screen and carry valuable information about the turbulent
ISM in the Faraday screen \citep{fletcher06}. Other features are
associated with real objects, like planetary nebulae
\citep{ransom08} or the photodissociation regions of molecular
clouds \citep{wolleben04}.

Little is known about the halo field in the Milky Way. The
synchrotron scale height of about 1.5~kpc indicates a scale height
of the total field of at least 6~kpc. The local regular Galactic
field, according to RM data from extragalactic sources, has no
significant vertical component towards the northern Galactic pole
and only a weak vertical component of $B_z\simeq0.3~\mu$G towards
the south \citep{mao10}.

\section{Galaxy clusters}

\begin{figure*}[t]
\vspace*{2mm}
\begin{minipage}[t]{6.3cm}
\includegraphics[bb = 39 38 572 625,width=6cm,clip=]{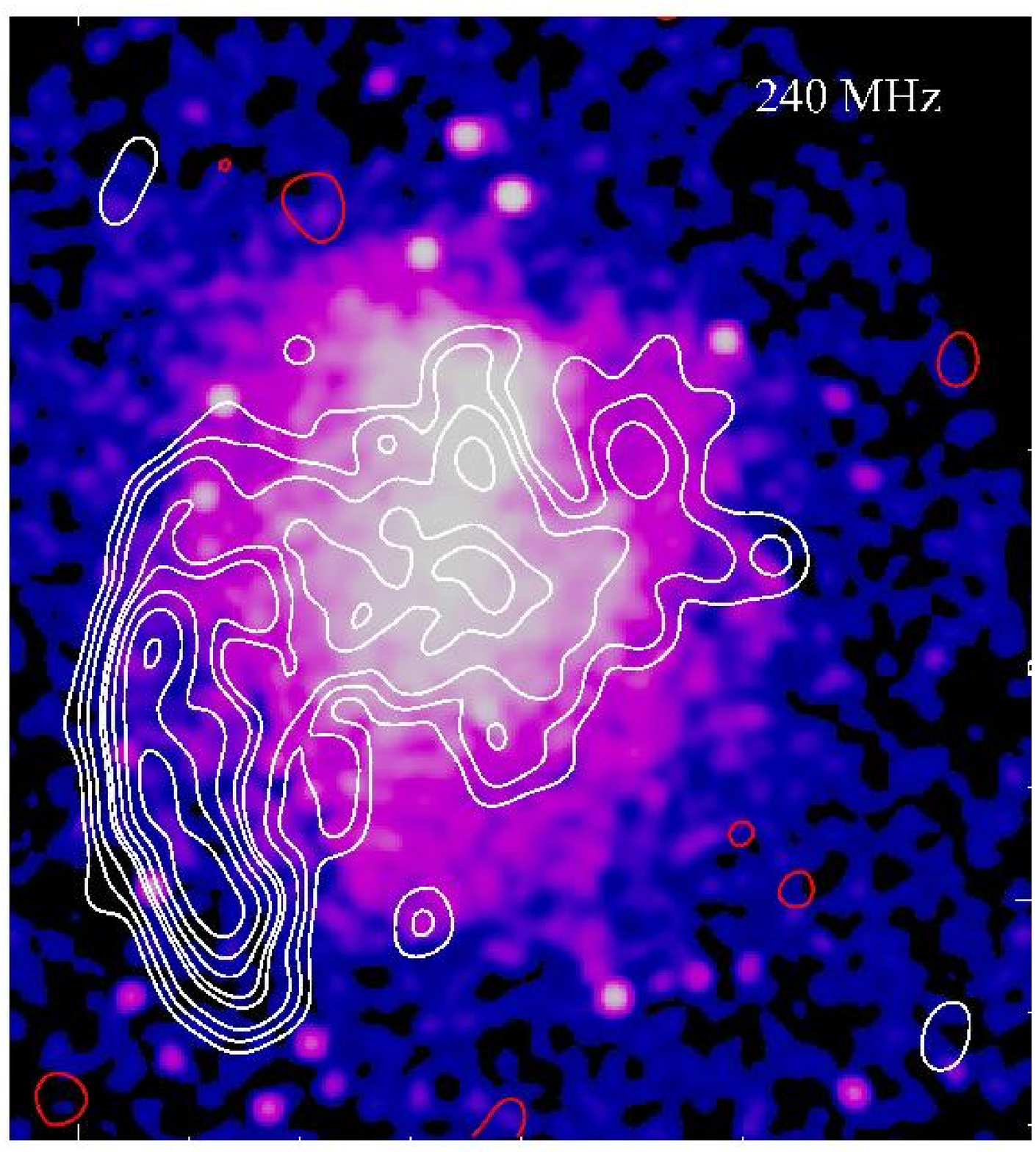}
\end{minipage}\hfill
\begin{minipage}[t]{8.5cm}
\includegraphics[bb = 32 32 545 378,width=8.3cm,clip=]{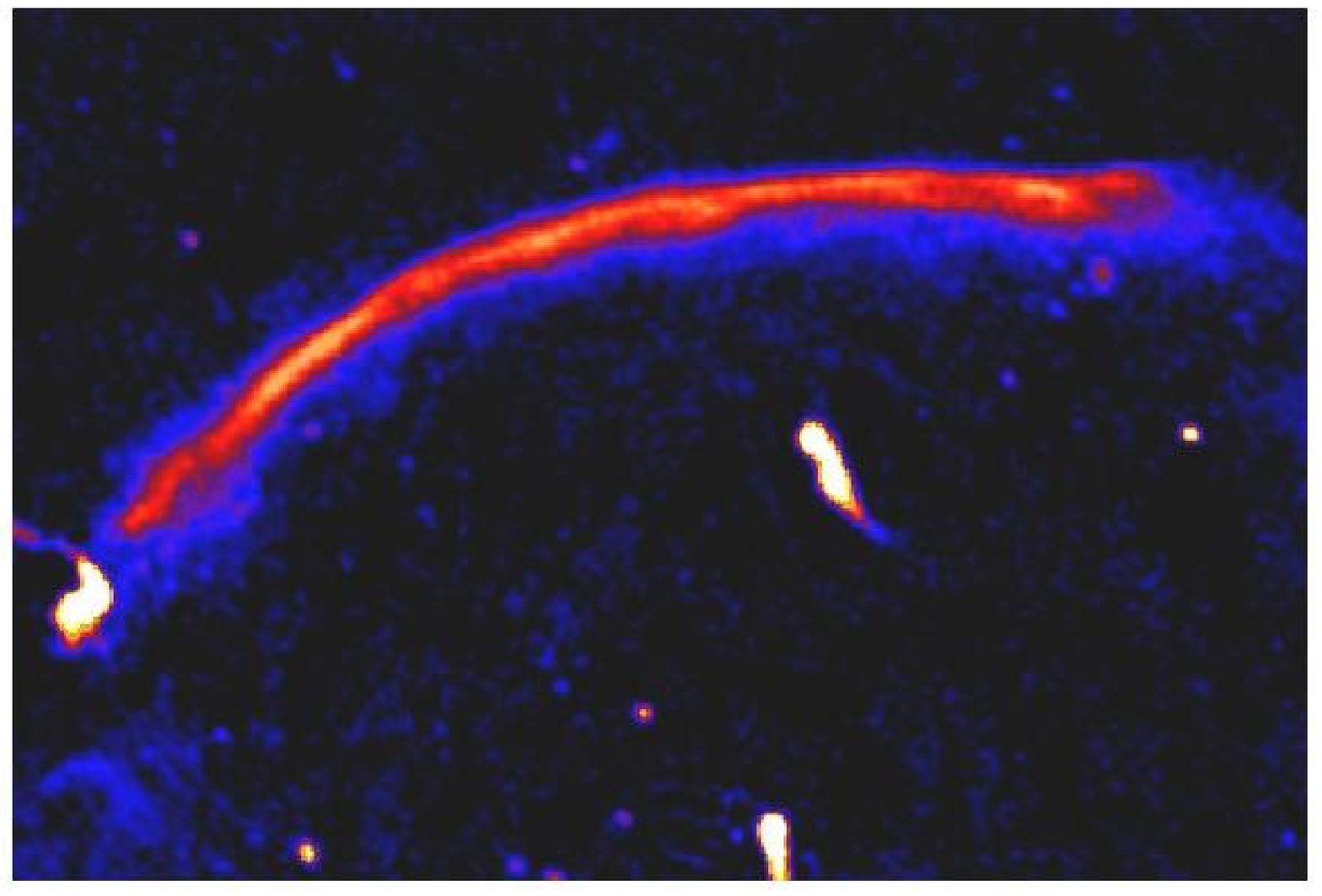}
\caption{{\bf Left:} Total radio intensity of the halo and relic
(left edge) of the galaxy cluster Abell~521, observed at 240~MHz
with the GMRT and smoothed to 35'', overlaid onto an X-ray image
from CHANDRA \citep{brunetti08}. {\bf Right:} Total radio intensity
of the relic (the ``sausage'') at the edge of the galaxy cluster
CIZA J2242.8+5301 at a redshift of 0.1921, observed at 610~MHz with
the GMRT at 4'' resolution \citep{weeren10}.} \label{fig:cluster}
\end{minipage}
\end{figure*}

Some fraction of galaxy clusters, mostly the massive and X-ray
bright ones, has diffuse radio emission \citep{cassano08}, emerging
from diffuse {\em halos}\ and steep-spectrum {\em relics}\ which can
best be observed at low frequencies (Fig.~\ref{fig:cluster}). The
diffuse halo emission is almost unpolarized and emerges from
turbulent intracluster magnetic fields. RMs towards background
sources show a vanishing mean value and a dispersion that decreases
with distance from the cluster center \citep{clarke01}. Relics can
emit highly polarized radio waves from anisotropic magnetic fields
generated by compression in merger shocks
\citep{ensslin98,govoni05,weeren10}. The record-holder is a relic in
a distant cluster of about 2~Mpc length and with a degree of
polarization of 50--60\% (Fig.~\ref{fig:cluster} right).

The asymmetric polarized intensities of galaxies observed in the
Virgo cluster \citep{vollmer07,wez07} (Fig.~\ref{fig:n5775} left)
can be interpreted as sweeping-up the intracluster magnetic field,
if the galactic field is shielded, similar to the interaction
between the solar wind and the Earth's magnetic field. The
asymmetries may indicate that the intracluster field has a radial
pattern, as predicted by the magneto-thermal instability
\citep{pfrommer10}. In this case, future high-sensitivity
observations should reveal weakly polarized emission from cluster
halos.

Equipartition strengths of the total magnetic field range from 0.1
to $1~\mu$G in halos \citep{govoni04}, and are higher in relics. On
the other hand, Faraday rotation data towards background sources
behind cluster halos reveals fields of a few $\mu$G strength
fluctuating on coherence scales of a few kpc \citep{govoni04} and
even fields of $40~\mu$G in the cores of cooling flow clusters
\citep{carilli02} where they may be dynamically important. The
reason for the difference in the field strength determinations is
still under discussion.

High-resolution RM maps of radio galaxies embedded in a cluster
allowed to derive the power spectra of the turbulent intracluster
magnetic fields which are of Kolmogorov type
\citep{bonafede10,vogt03,vogt05}.

\section{Intergalactic magnetic fields}

Magnetic fields in the intergalactic medium (IGM) are of fundamental
importance for cosmology \citep{widrow02}. Their role as the likely
seed field for galaxies and clusters and their possible relation to
structure formation in the early Universe place considerable
importance on its discovery. Various generation mechanisms have been
suggested. The field could be produced via the Weibel instability at
structure formation shocks \citep{medvedev04}. Another possibility
is the injection from galactic black holes (AGNs) and outflows from
starburst galaxies \citep{voelk00}. In each case the field is
subsequently amplified by compression and large-scale shear flows
\citep{brueggen05}. Ryu et al. \citep{ryu08} have argued that highly
efficient amplification is possible via MHD turbulence, with the
source of the turbulent energy being the structure formation shocks
themselves. Estimates of the strength of the turbulent field in
filaments obtained from MHD simulations with a primordial seed field
typically range between 0.1~$\mu$G and 0.01~$\mu$G, while regular
fields are weaker.

Until recently, there has been no unambiguous detection of a general
magnetic field in the IGM. In an intergalactic region of about
$2^{\circ}$ extent west of the Coma Cluster, containing a group of
radio galaxies, enhanced synchrotron emission may indicate an
equipartition total field strength of $0.2-0.4~\mu$G
\citep{kronberg07}. Xu et al. \citep{xu06} observed an excess of
rotation measures (RM) towards two super-clusters which may indicate
regular magnetic fields of $<0.3~\mu$G on scales of order 500~kpc.
Gamma-ray halos around active galactic nuclei can be used to measure
the strength of the intergalactic field \citep{ando10}, but the
observed halos could also be instrumental effects. Lee et al.
\citep{lee09} found indications for a statistical correlation at the
4$\sigma$ level of the RMs of background sources with the galaxy
density field which may correspond to an intergalactic field of
about 30~nG strength and about 1~Mpc coherence length. However, the
current data are probably insufficient to constrain the amplitude
and distribution of large-scale intergalactic fields
\citep{stasyszyn10}.

Progress came with the FERMI satellite. The non-observation of GeV
$\gamma$-ray emission from the electromagnetic cascade in the IGM
initiated by TeV $\gamma$-rays from blazars observed with HESS
yields {\em a lower limit of the IGM magnetic field strength}\ of
$3~10^{-16}$~G \citep{neronov10}. This IGM must be all-pervasive,
with a filling factor of at least 60\% \citep{dolag11}.

\section{Prospects}

\begin{figure*}[t]
\includegraphics[bb = 51 51 519 332,width=9.5cm,clip=]{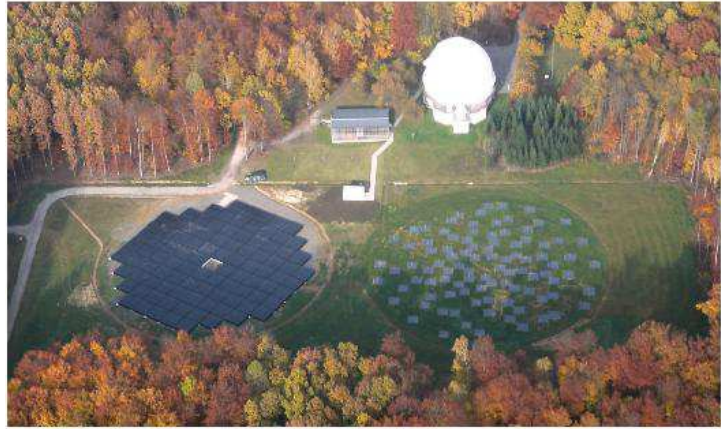}
\caption{LOFAR station in Tautenburg/Germany (Copyright: Michael
Pluto, TLS).} \label{fig:lofar}
\end{figure*}

Next-generation radio telescopes will widen the range of observable
magnetic phenomena. At low frequencies, synchrotron emission can be
observed from aging electrons far away from their places of origin.
Low frequencies are also ideal to search for small Faraday rotation
measures from weak interstellar and intergalactic fields
\citep{beck09} and in steep-spectrum cluster relics
\citep{brunetti08}. The recently completed {\em Low Frequency
Array}\ (LOFAR) (Fig.~\ref{fig:lofar}), followed by the {\em
Murchison Widefield Array}\ (MWA) and the {\em Long Wavelength
Array}\ (LWA) (both under construction), are suitable instruments to
search for weak magnetic fields in outer galaxy disks, galaxy halos
and cluster halos.

LOFAR will detect all pulsars within 2~kpc of the sun and discover
about 1000 new nearby pulsars, especially at high latitudes
\citep{leeuwen10}. Most of these are expected to emit strong,
linearly polarized signals at low frequencies. This will allows us
to measure their RMs and to derive an unprecedented picture of the
magnetic field near to the sun.

\begin{figure*}[t]
\vspace*{2mm}
\begin{minipage}[t]{7.5cm}
\begin{center}
\includegraphics[bb = 52 52 519 339,width=7cm,clip=]{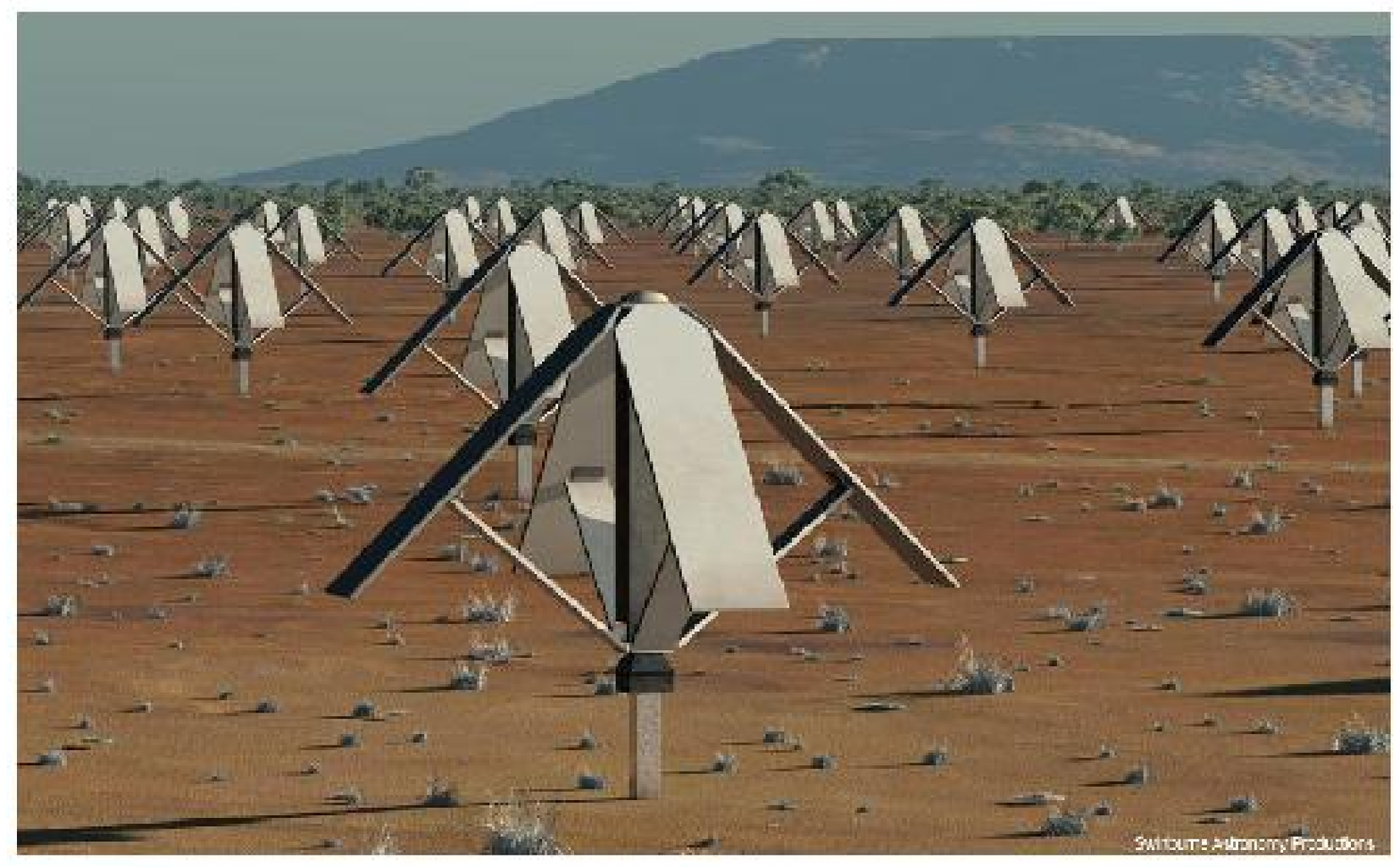}
\end{center}
\end{minipage}\hfill
\begin{minipage}[t]{7.5cm}
\begin{center}
\includegraphics[bb = 52 52 519 339,width=7cm,clip=]{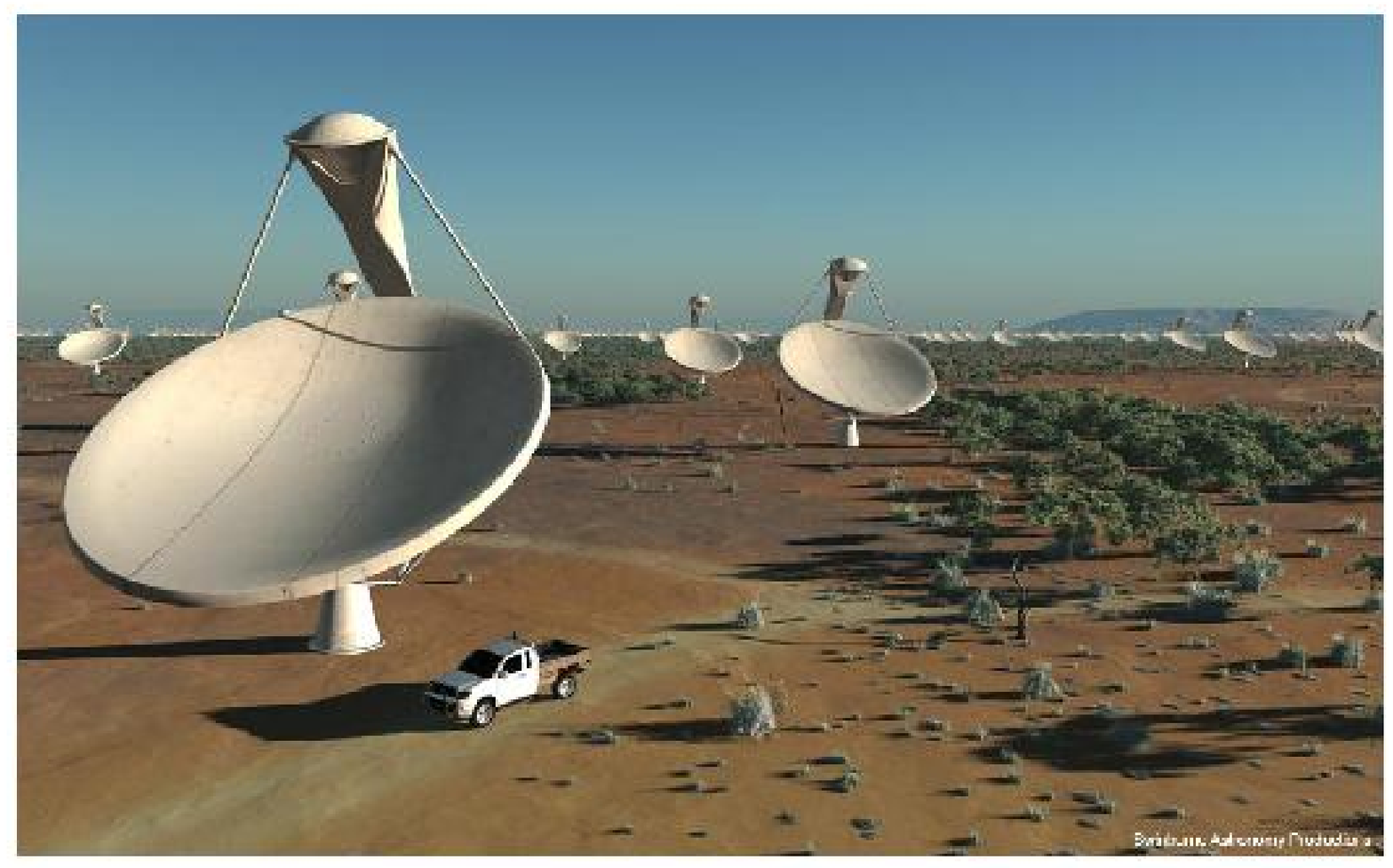}
\caption{Planned SKA configurations: dipoles for low frequencies
(70--450~MHz) (left) and dishes for high frequencies
(450~MHz--3~GHz) (Copyright: Swinburne Astronomy Productions and SKA
Project Development Office).} \label{fig:ska_conf}
\end{center}
\end{minipage}
\end{figure*}

Deep high-resolution observations at high frequencies, where Faraday
effects are small, require a major increase in sensitivity of
continuum observations, to be achieved by the {\em Extended
Very Large Array}\ (EVLA) and the planned {\em Square Kilometre
Array}\ (SKA) \citep{beck10} (Fig.~\ref{fig:ska_conf}). The detailed
structure of the magnetic fields in the ISM of galaxies, in galaxy
halos, cluster halos and cluster relics will be observed.
The magnetic power spectra can be measured.
Direct insight into the interaction between gas and magnetic fields
in these objects will become possible. The SKA will also allow to
measure the Zeeman effect of weak magnetic fields in the Milky Way
and in nearby galaxies.

\begin{figure*}[t]
\vspace*{2mm}
\begin{minipage}[t]{8.3cm}
\begin{center}
\includegraphics[bb = 37 22 531 387,width=8.2cm,clip=]{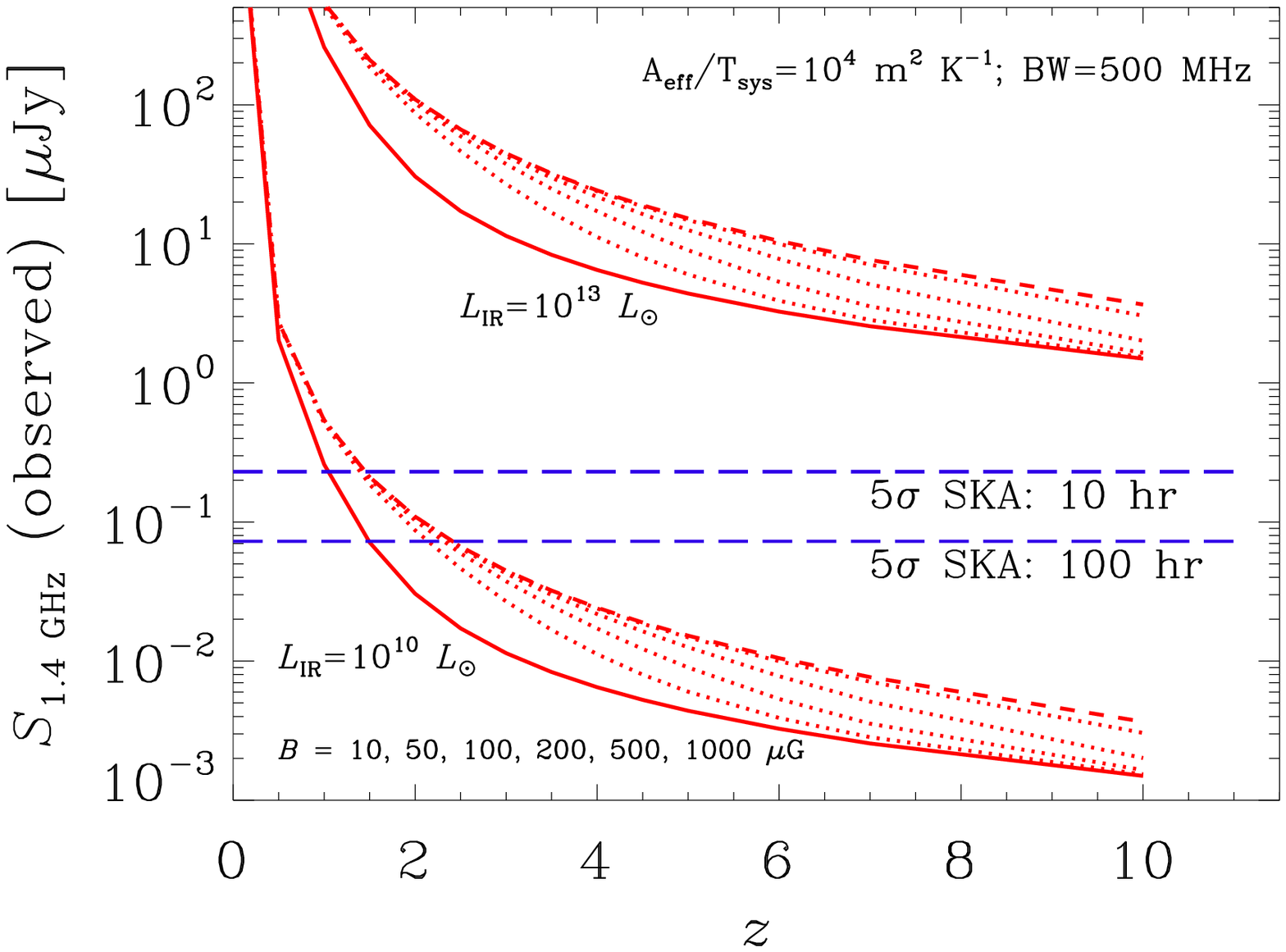}
\end{center}
\end{minipage}\hfill
\begin{minipage}[t]{6.5cm}
\begin{center}
\includegraphics[bb = 31 31 325 325,width=6.3cm,clip=]{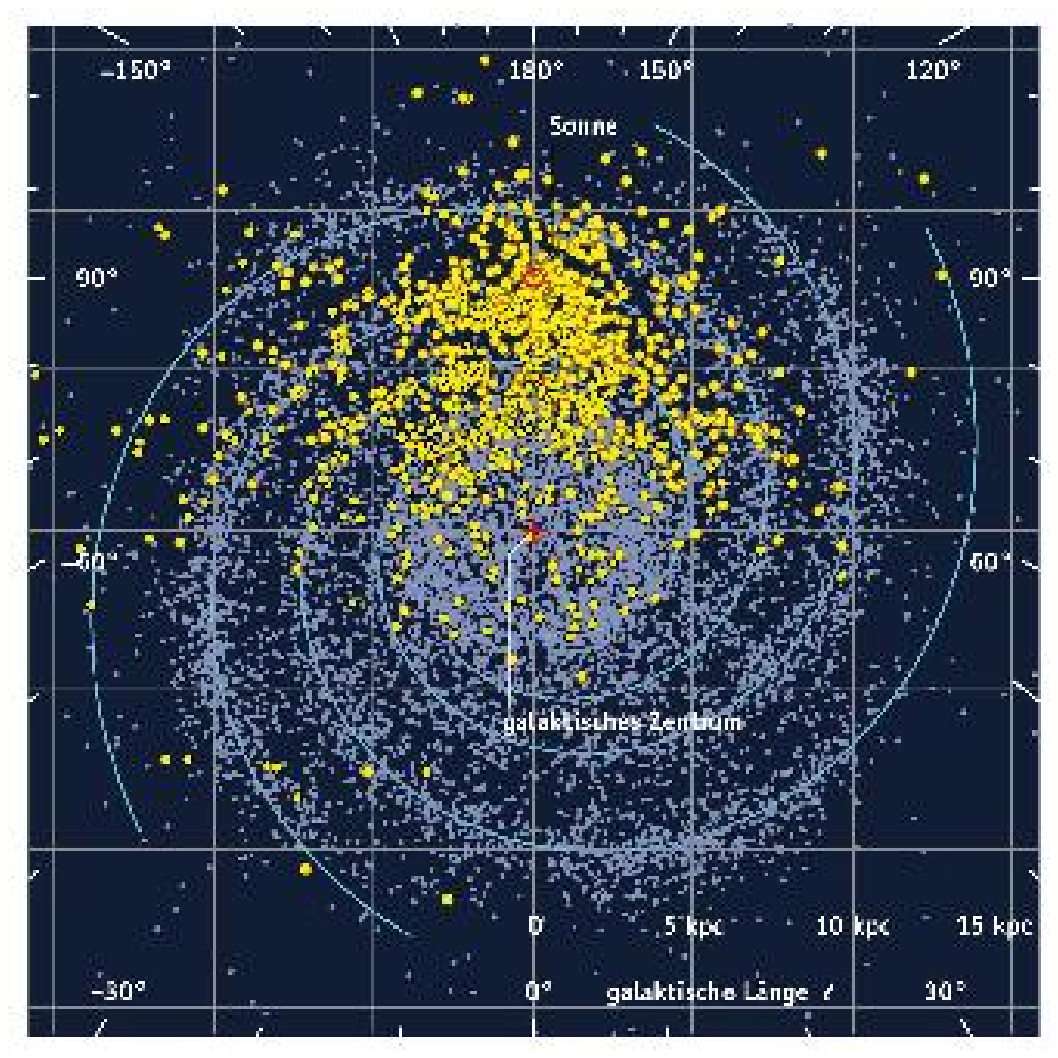}
\caption{{\bf Left:} Total synchrotron emission at 1.4~GHz as a
function of redshift $z$ and magnetic field strength $B$, and the
$5\sigma$ detection limits for 10~h and 100~h integration time with
the SKA \citep{murphy09}. {\bf Right:} Simulation of about 20,000
pulsars (blue) in the Milky Way that will be detected with the SKA,
compared to about 2000 pulsars known today (yellow). Graphics:
\textit{Sterne und Weltraum} (from Cordes, priv. comm).}
\label{fig:ska}
\end{center}
\end{minipage}
\end{figure*}

Detection of polarized emission from distant, unresolved galaxies
will reveal large-scale ordered fields \citep{stil09}, to
be compared with the predictions of dynamo theory
\citep{arshakian09}. The SKA at 1.4~GHz will detect Milky-Way type
galaxies at $z\le1.5$ (Fig.~\ref{fig:ska} left) and their polarized
emission at $z\le0.5$ (assuming 10\% polarization). Bright starburst
galaxies can be observed at larger redshifts, but are not expected
to host ordered fields. Cluster relics are also
detectable at large redshifts through their integrated polarized
emission.

Unpolarized synchrotron emission, signature of turbulent magnetic
fields, can be detected with the SKA in starburst galaxies out to
large redshifts, depending on luminosity and magnetic field strength
(Fig.~\ref{fig:ska} left), and for cluster halos. However, for
fields weaker than $3.25~\mu$G $(1+z)^2$, energy loss of cosmic-ray
electrons is dominated by the inverse Compton effect with CMB
photons, so that the energy appears mostly in X-rays, not in the
radio range. On the other hand, for strong fields the energy range
of the electrons emitting at a 1.4~GHz drops to low energies, where
ionization and bremsstrahlung losses become dominant
\citep{murphy09}. In summary, the mere detection of synchrotron
emission at high redshifts will constrain the range of allowed
magnetic field strengths.

If polarized emission from galaxies, cluster halos or cluster relics
is too weak to be detected, the method of {\em RM grids}\ towards
background QSOs can still be applied and allows us to determine the
field strength and pattern in an intervening galaxy. This method can
be applied to distances of young QSOs ($z\simeq5$). Regular fields
of several $\mu$G strength were already detected in distant galaxies
\citep{bernet08,kronberg92,kronberg08}. Mean-field dynamo theory
predicts RMs from evolving regular fields with increasing coherence
scale at $z\le3$ \citep{arshakian09}. (Note that the observed RM
values are reduced by the redshift dilution factor of $(1+z)^{-2}$.)
A reliable model for the field structure of nearby galaxies, cluster
halos and cluster relics needs RM values from a large number of
polarized background sources, hence large sensitivity and/or high
survey speed \citep{krause+09}.

The {\em POSSUM survey}\ at 1.4~GHz with the {\em Australia SKA
Pathfinder}\ (ASKAP) telescope (under construction) with 30~deg$^2$
field of view will measure about 100 RMs of extragalactic sources
per square degree within 10~h integration time, about 100~times
denser than in Fig.~\ref{fig:Galaxy} (left). Similarly long
integrations with the EVLA and with MeerKAT will show about 5~times
more RMs, but their fields of view are smaller.

The {\em SKA Magnetism Key Science Project}\ plans to observe a
wide-field survey (at least $10^4$~deg$^2$) around 1~GHz with 1~h
integration per field which will detect sources of $0.5-1~\mu$Jy
flux density and measure at least 1500 RMs deg$^{-2}$. This will
contain at least $2~10^7$ RMs from compact polarized extragalactic
sources at a mean spacing of $\simeq90''$ \citep{gaensler04}. This
survey will also be used to model the structure and strength of the
magnetic fields in the Milky Way, in intervening galaxies and
clusters and in the intergalactic medium \citep{beck04}. The SKA
pulsar survey will find about 20,000 new pulsars which will mostly
be polarized and reveal RMs (Fig.~\ref{fig:ska} right), suited to
map the Milky Way's magnetic field with high precision. More than
10,000 RM values are expected in the area of M~31 and will allow the
detailed reconstruction of the 3-D field structure. Simple patterns
of regular fields can be recognized out to distances of about
100~Mpc \citep{stepanov08} where the polarized flux is far too low
to be mapped. The evolution of field strength in cluster halos can
be measured by the RM grid to redshifts of about 1
\citep{krause+09}.

If the filaments of the local Cosmic Web outside clusters contain a
magnetic field \citep{ryu08}, possibly enhanced by IGM shocks, we
hope to detect this field by direct observation of its total
synchrotron emission \citep{keshet04} and possibly its polarization,
or by Faraday rotation towards background sources. For fields of
$\approx 10^{-8}-10^{-7}$~G with 1~Mpc coherence length and
$n_e\approx 10^{-5}$~cm$^{-3}$ electron density, Faraday rotation
measures between 0.1 and 1~rad m$^{-2}$ are expected which will be
hard to detect even with LOFAR. More promising is a statistical
analysis like the measurement of the power spectrum of the magnetic
field of the Cosmic Web \citep{kolatt98} or the cross-correlation
with other large-scale structure indicators like the galaxy density
field \citep{stasyszyn10}.

If an overall IGM field with a coherence length of a few Mpc existed
in the early Universe and its strength varied proportional to
$(1+z)^2$, its signature may become evident at redshifts of $z>3$.
Averaging over a large number of RMs is required to unravel the IGM
signal. The goal is to detect an IGM magnetic field of 0.1~nG, which
needs an RM density of $\approx1000$ sources deg$^{-2}$
\citep{kolatt98}, achievable with the SKA. Detection of a general
IGM field, or placing stringent upper limits on it, will provide
powerful observational constraints on the origin of cosmic
magnetism.

\end{document}